\documentclass[12pt, journal,draftclsnofoot, onecolumn]{IEEEtran}
\usepackage{cite}
\usepackage{ulem}
\usepackage{bm}
\usepackage{graphicx}
\usepackage{amsmath,bm}
\usepackage{array}
\usepackage{mdwmath}
\usepackage{amssymb}
\usepackage{mdwtab}
\usepackage{stfloats}
\usepackage[tight,footnotesize]{subfigure}
\usepackage{amsmath,amsthm}
\usepackage{threeparttable}
\usepackage{color}
\usepackage{ulem}
\usepackage[ruled,linesnumbered]{algorithm2e}
\usepackage{multirow}
\usepackage{float}

\makeatletter
\newcolumntype{"}{@{\hskip\tabcolsep\vrule width 3pt\hskip\tabcolsep}}
\makeatother

\newtheorem{theorem}{\it Theorem}

\newtheorem{lemma}{\it  Lemma}

\begin{document}
	\title{ Adaptive Modulation for Wobbling UAV Air-to-Ground Links in Millimeter-wave Bands}
	\author{Songjiang Yang,~\textit{Graduate~Student~Member,~IEEE}, Zitian Zhang, \\ Jiliang Zhang,~\textit{Senior~Member,~IEEE}, Xiaoli Chu,~\textit{Senior~Member,~IEEE}, \\Jie Zhang,~\textit{Senior~Member,~IEEE}
	\thanks{Songjiang Yang and Xiaoli Chu are with the Department of Electronic and Electrical Engineering, the University of Sheffield, Sheffield, S10 2TN, UK, (syang16@sheffield.ac.uk, jiliang.zhang@sheffield.ac.uk).

	Zitian Zhang is with Ranplan Wireless Network Design Ltd., Cambridge, CB23 3UY, UK.		
		
	Jiliang Zhang and Jie Zhang is with the Department of Electronic and Electrical Engineering, the University of Sheffield, Sheffield, S10 2TN, UK, and also with Ranplan Wireless Network Design Ltd., Cambridge, CB23 3UY, UK.}}
	\markboth{IEEE TWC,~Vol.~X, No.~X, Month~20XX}
	{Shell \MakeLowercase{\textit{et al.}}: Bare Demo of IEEEtran.cls for Journals}
	\maketitle

	\begin{abstract}	
	The emerging millimeter-wave (mm-wave) unmanned aerial vehicle (UAV) air-to-ground (A2G) communications are facing the Doppler effect problem that arises from the inevitable wobbling of the UAV.  
	The fast time-varying channel for UAV A2G communications may lead to the outdated channel state information (CSI) from the channel estimation.
	In this paper, we introduce two detectors to demodulate the received signal and get the instantaneous bit error probability (BEP) of a mm-wave UAV A2G link under imperfect CSI. 
	Based on the designed detectors, we propose an adaptive modulation scheme to maximize the average transmission rate under imperfect CSI by optimizing the data transmission time subject to the maximum tolerable BEP.
	A power control policy is in conjunction with adaptive modulation to minimize the transmission power while maintaining both the BEP under the threshold and the maximized average transmission rate.
	Numerical results show that the proposed adaptive modulation scheme in conjunction with the power control policy could maximize the temporally averaged transmission rate, while saves as much as 50\% energy.
	\end{abstract}

	
	\section{Introduction}
	\normalem
	Unmanned aerial vehicle (UAV) air-to-ground (A2G) communications have been widely investigated for providing flexible coverage and capacity enhancements, where UAVs carry aerial base stations (BSs) or mobile relay nodes as part of 5G cellular systems \cite{Zeng2016COM, Hou2020TVT}. 
	UAV A2G links have also been designed to operate in millimeter-wave (mm-wave)  bands to support high data rates \cite{Xiao2016COM}. 
	
	While hovering in the air, UAVs will be inevitably wobbling due to various environmental and mechanical issues, such as wind gusts, bad weather, and high vibration frequency of their propellers and rotors \cite{Banagar2020TVT}. 
	Such UAV wobbling leads to the Doppler effect on the mm-wave UAV A2G link \cite{Xiao2016COM, Zhang2019WCZ}, making the wireless channel unstable and unpredictable \cite{Banagar2020TVT}.
	In addition, under the size, weight, and battery capacity restrictions of UAVs, the power consumption has become a bottleneck for the UAV-carried aerial BSs or mobile relay nodes \cite{Chen2013TVT}.
	
	\subsection{Related Works}
	There has been a surge of recent interest in the impact of UAV wobbling on mm-wave UAV A2G communications \cite{Banagar2020TVT, Dabiri2020TWC, YangArxiv}.
	In \cite{Banagar2020TVT}, the authors studied the impact of UAV wobbling on the channel coherence time between the UAV and ground user equipment.
	In \cite{Dabiri2020TWC}, the authors studied the antenna mismatch between the transmitter and the receiver of a mm-wave UAV A2G link caused by the UAV wobbling.
	In \cite{YangArxiv}, the authors derived the Doppler power spectrum density in mm-wave bands under the UAV wobbling.
	Although the channel state information (CSI) of a UAV A2G link can be estimated with a low mean square error under the Doppler effect by exploiting some known channel characteristics \cite{Cirpan1999TSP, Yang2012TB}, the channel estimation for the time-varying channel has to be updated sufficiently frequently; otherwise the outdated CSI will increase the bit error probability (BEP) and reduce the throughput of the UAV A2G link \cite{Li2009CL}. 	
	
	Adaptive modulation technique has been widely used in mobile communication systems to improve the spectral efficiency and throughput by adapting the modulation order to the time-varying channel \cite{Svensson2007PIEE}. 
	In \cite{Yu2009TVT}, assuming perfect CSI for an adaptively modulated multiple-input-multiple-output (MIMO) system, the multiple modulation order regions were divided by comparing the signal-to-noise ratio (SNR) at the receiver with a predefined switching threshold to maximize the spectral efficiency subject to the BEP and power constraints.
	In \cite{Liu2016CL}, the authors proposed a signal-to-interference-plus-noise-ratio (SINR) estimation algorithm under imperfect CSI and used the estimated instantaneous SINR to decide the modulation order to improve the spectral efficiency.
	In \cite{Akin2015TComm}, the authors analyzed the impact of imperfect CSI on an adaptively modulated terrestrial communication system by using discrete correlation coefficients between the predicted and the true SNR at the receiver.	
	On the contrary, for a UAV A2G link, a continuous temporal correlation function of the channel impulse response is more suitable to describe the CSI than discrete correlation coefficients because the hovering UAV is not completely stationary causing the time-varying Doppler effects on the A2G channel \cite{ Banagar2020TVT}.  
	We note that the trade-off between the channel estimation time and the data transmission time has not been sufficiently studied to maximize the average transmission rate of an adaptive modulation scheme under imperfect CSI. 
	
	Although adaptive modulation schemes can be used to improve the spectral efficiency of an A2G link, they need to be supported by a power control policy to maintain the energy efficiency because using the constant transmission power while reducing the modulation order will waste the energy of the UAV \cite{Zarakovitis2016JSAC}.
	In \cite{Chakareski2019TGCN}, the authors proposed power allocation mechanisms for fixed BSs and steady UAVs to maximize the system sum rate or minimize the system  power consumption of a UAV-assisted mm-wave heterogeneous cellular network.
	In \cite{Huang2019TComm}, the authors maximized the average achievable rate of a steady UAV A2G link, which shared the spectrum with terrestrial wireless communication links, by optimizing the UAV's 3D trajectory and transmission power.
	We note that the combination of the adaptive modulation and the power control for wobbling mm-wave UAV A2G links to maintain the maximum achievable transmission rate has not been studied in the published literature.
	\subsection{Contributions}		 
	In this paper, we propose a novel adaptive modulation scheme in conjunction with a power control policy to maximize the average transmission rate and minimize the instantaneous transmission power of a mm-wave UAV A2G link simultaneously while considering imperfect CSI.
	More specifically, we derive the optimum transmission time of the adaptive modulation scheme that maximizes the average transmission rate of the wobbling mm-wave UAV A2G link subject to the maximum tolerable BEP.
	The power control policy minimizes the instantaneous transmission power for the adaptive modulation scheme, while maintaining the maximized average transmission rate subject to the BEP threshold. 
	The major contributions of this paper are summarized as follows:
	\begin{itemize}
	\item
	The imperfect CSI is modeled considered a continuous temporal autocorrelation function (ACF) of the mm-wave UAV A2G channel impulse response.
	Then the relationship between BEP and temporal ACF can be analytical derived, which could be used to adjust the modulation order in the adaptive modulation scheme accordingly. 
	\item 
	Two detectors are introduced for mm-wave UAV A2G links under imperfect CSI, i.e., the maximum likelihood detector and the sub-optimum detector, which both use the continuous temporal ACF as the reference to demodulate the received signal and get the instantaneous BEP.
	To speed up the calculation of the instantaneous BEP, the sub-optimum detector ignores the power differences among symbols in a constellation diagram, leading to a computational complexity lower than that of the maximum likelihood detector. 
	We derive the closed-form union upper bound (UUB) on the BEP for the sub-optimum detector under imperfect CSI and verify its accuracy by comparing it with the Monte-Carlo simulations.
	\item 
	The adaptive modulation scheme is proposed based on the designed detectors to  maximize the average
	 transmission rate of the mm-wave UAV A2G link under imperfect CSI. 
	 The data transmission time and adapting the modulation order (among \textit{M}-ary phase shift keying (PSK) or \textit{M}-ary quadrature amplitude modulation (QAM)) is optimized subject to a BEP threshold, constant transmission power, and a channel estimation time constraint.
	\item
	 {\color{blue}A new power control policy is designed to minimize the instantaneous transmission power at the receiver subject to the BEP threshold and the transmission power constraint, while maintaining the maximum instantaneous transmission rate. 
	 The signal-space concept BEP approximation for \textit{M}-ary PSK and the Netwon-Raphson method for \textit{M}-ary QAM are used to compute the minimized instantaneous transmission power. }
	\item
	Numerical results show that the proposed sub-optimum detector with a lower computational complexity achieves a similar demodulation accuracy as that of the maximum likelihood detector. 
	The proposed optimum adaptive modulation scheme in conjunction with the power control policy is able to achieve the maximum average transmission rate of the mm-wave UAV A2G link under imperfect CSI, while keeping the minimum transmission power consumption subject to the BEP threshold.
	\end{itemize}  

	The rest of this paper is organized as follows. 
	In Section II, the system model of a mm-wave UAV A2G link with adaptive modulation under imperfect CSI are presented. 
	In Section III, the maximization of the average transmission rate is formulated and solved.
	In Section IV, the minimization of the transmission power is formulated and solved.
	In Section V, numerical results are provided to evaluate the performance of the optimum adaptive modulation scheme and the power control policy under imperfect CSI. 
	Finally, in Section VI, our main conclusions are drawn.
	\section{System Model}
	\begin{figure}[t]
		\centering
		\includegraphics[width=0.6\linewidth]{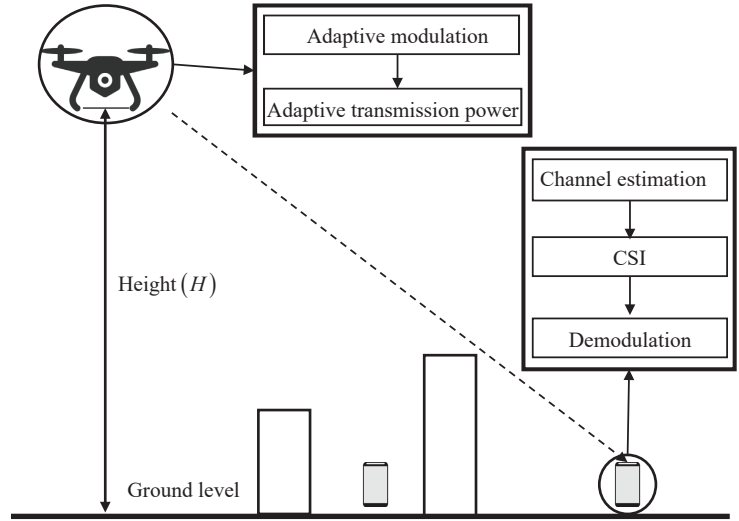}
		\caption{The mm-wave UAV A2G link.}
		\label{fig:uavsystemmodel}
	\end{figure}
	We consider a wobbling mm-wave UAV A2G link, as illustrated in Fig. \ref{fig:uavsystemmodel}, where the ground node is equipped with $N_\mathrm{R}$ antennas, while the UAV has a single antenna due to the weight and size constraints \cite{Ponce2021ICC}. 
	We assume that the UAV hovers at a height $ H $ above the ground and its horizontal location is $ \left(0, 0 \right)$. The horizontal location of the ground node is denoted by $ \mathbf{w} \in \mathbb{R}^{2 \times1} $.
	Thus, the distance between the UAV and the ground node is given by $ d=\sqrt{H^2+\left\|\mathbf{w}\right\|^2} $. 
	Since the probability of a line-of-sight (LoS) link between the UAV and the ground node is high and the dominant propagation mechanism is free space transmission \cite{Zeng2019PIEEE}, the path loss between the UAV and the ground node can be expressed as \cite{Rappaport2002}
	\begin{equation}\label{eq:pathloss}
		P_\mathrm{L}=\left( \frac{4\pi d f}{c}\right) ^2,
	\end{equation}
	where $ f$ is the signal frequency and $c $ is the speed of light.
	  
\begin{figure}[t]
		\centering
		\includegraphics[width=0.5\linewidth]{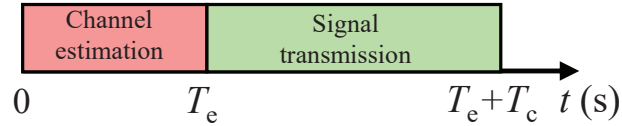}
		\caption{The schematic diagram of the transmission frame structure.}
		\label{fig:Framestr}
	\end{figure}

	Each transmission frame is composed of a fixed channel estimation period $T_\mathrm{e}$ (seconds) followed by a variable signal transmission period $T_\mathrm{c}$ (seconds), which is long enough to allow the transmission of at least one symbol. 
	In the channel estimation period, the UAV transmits mutually orthogonal pilot sequences to the $N_\mathrm{R}$ receiving antennas at the ground node, where the CSI of the A2G channel is estimated for the current transmission frame.
	{\color{blue} Based on the available channel knowledge, the ground node optimizes the transmission policy to increase the throughput and reduce the transmission power after the channel estimation period. 
	Then, we assume that the transmission policy is fed back to the UAV without errors and delay.
	In the signal transmission period, the UAV will transmit data signal based on the transmission policy and the ground node receiver will use the estimated CSI to demodulate the received signal.
	The frame structure schematic diagram is shown in Fig \ref{fig:Framestr}.}
	
	{\color{blue} Without loss of generality, we assume that the current transmission frame starts at time $ t $ = 0. 
	 The average transmission rate in a transmission frame is given by
	\begin{equation}\label{eq:Ravedef}
		R_{\rm{ave}}\left( T_{\rm{c}} \right) = \frac{1}{T_{\mathrm{e}}+T_{\mathrm{c}}}\int\limits_{T_{\mathrm{e}}}^{T_{\mathrm{e}}+T_{\rm{c}}}	{R\left( t \right)\mathrm{d} t}, 
	\end{equation}
	where $R\left( t \right) $ is the transmission rate (bit/symbol) at time $ t $.}
		
	The UAV communication power consumption is composed of the communication circuitry power and the transmission power.
	The circuitry power consumption is typically negligible compared to the transmission power and is thus ignored hereafter for analytical simplicity \cite{Chen2013TVT}.
	\subsection{Time-varying Channel Impulse Response}
	At the end of the channel estimation period $ T_\mathrm{e} $, the ground node receiver obtains the estimated channel impulse response $ \mathbf{\hat{h}}(T_\mathrm{e})$, which is an $ N_\mathrm{R} $-by-1 vector and will be used by the ground node detector for signal demodulation during the entire signal transmission period. 
	If the channel estimation at $ T_\mathrm{e}$ is perfect, then the estimated channel impulse response is equal to the actual channel impulse response, i.e., $\mathbf{\hat{h}}(T_\mathrm{e})= \mathbf{h}(T_\mathrm{e}) $. 
	However, due to UAV wobbling, the actual channel impulse response $ \mathbf{h}( t)  $  in the signal transmission period may become different from $\mathbf{h}(T_\mathrm{e})$ and is given by \cite{Goldsmith2005},
	\begin{equation}
		\label{equ:ht}
		\mathbf{h}\left( t \right) = \mathbf{h}\left( T_\mathrm{e} \right)C\left(T_\mathrm{e},t\right) + \mathbf{h_{\mathrm{rd}}}\sqrt {1 - {|C\left( T_\mathrm{e},t\right) |^2}},
	\end{equation}
	{\color{blue} where $C\left( T_\mathrm{e},t\right) = \mathrm{E}\left[\mathbf{h}(T_\mathrm{e})\mathbf{h}^*(t) \right] $ is the temporal ACF of the mm-wave UAV A2G channel impulse response,} and $\mathbf{h_\mathrm{rd}}$ is a $N_\mathrm{R}$-by-1 vector consisting of random elements each following an independent and identical Gaussian distribution $ \mathcal{CN}\left( \mathbf{0},\mathbf{I} \right)  $. Hence, the channel estimation $ \hat{\mathbf{h}}(T_\mathrm{e}) $, even though perfect at $ T_\mathrm{e} $, becomes imperfect CSI at time $ t $ in the signal transmission period. 
	
	The received signal at the ground node receiver at time $ t $ in the signal transmission period is given by
	\begin{equation}
		\label{equ:received}
		\mathbf{y}\left(t \right) = \sqrt{\gamma}  \mathbf{h}\left(t \right)s + \mathbf{n},
	\end{equation}
	where $ \gamma = \frac{P_\mathrm{T}}{P_\mathrm{L}N_0}$ is the average SNR \cite{Fu2016TVT},  $P_\mathrm{T}$ is the UAV transmission power,  $N_0 $ is the power of thermal noise, $ s $ is the transmitted symbol, and $\mathbf{n} \sim  \mathcal{CN}\left( 0, \mathbf{I}_{N_\mathrm{R}}\right)  $ is the additive white Gaussian noise (AWGN).
	\subsection{Adaptive Modulation}
	We consider an adaptive modulation scheme employing the \textit{M}-ary PSK or \textit{M}-ary QAM, $ M = \left\lbrace 2, 4, 8, 16, \dots\right\rbrace $, as the component modulations.
	The modulation order $ M $ is adaptively changed according to how the instantaneous BEP $ \beta_{i} $ is compared to a BEP threshold $ \beta_{\mathrm{th}} $ \cite{Agiwal2016TUT}.		
	Denoting the maximum achievable modulation order that maintains the instantaneous BEP below the BEP threshold under perfect CSI by $ M_{\max} $, then the maximum achievable transmission rate is given by $ R_{\max} = \log_2M_{\max}$. 
	
	The adaptive transmission rate at time $ t $ in the current transmission frame under imperfect CSI is given by
	\begin{equation}
		\label{equ:rate_ac}
		R\left( t \right) = \left\{ \begin{array}{ll} 
			n , &C_{n} <  C\left(T_\mathrm{e},t \right)  \le  C_{n+1},\\
			0, &  C\left(T_\mathrm{e},t \right) \le C_1,
		\end{array} \right.	
	\end{equation}
	where $ n \in \left\lbrace  1, \dots, R_{\max} \right\rbrace $ is the transmission rate when the value of the temporal ACF $ C\left(T_\mathrm{e},t \right) $ is between $ C_n$ and $C_{n+1} $,  $ C_n \in \left\{ {C_1}, \cdots, {C_{R_\mathrm{max}}} \right\}$ stands for the minimum required temporal ACF for the instantaneous BEP to be kept below the BEP threshold when the transmission rate is $ n $, $ C_{R_{\max}+1}=1 $.
	The value of $ C_n $ can be computed by solving the equation BEP$\left[C\left(T_\mathrm{e},t \right) \right]=\beta_{\mathrm{th}}$ for $ C\left(T_\mathrm{e},t \right) $ and $2^n$-ary PSK or QAM using the dichotomy method, where the expression of BEP$\left[C\left(T_\mathrm{e},t \right) \right]$ will be given in \eqref{equ:ABEP_ana} in next subsection.
	{\color{blue} According to \cite[Fig. 3]{Banagar2020TVT}, the temporal ACF $ C\left(T_\mathrm{e},t \right) $ under UAV wobbling monotonically decreases with $ t $, hence there is a one-to-one match between  the temporal ACF $ C_n $ and the corresponding transmission time $ t_n $, i.e., $ C_n = C\left(T_\mathrm{e}, t_n \right)  $, and $ C_{n+1} > C_n $ for $ t_{n+1}<t_n $.} 
	Thus, the adaptive transmission rate at time $ t$ in the current transmission frame under imperfect CSI can be rewritten as
	\begin{equation}
		\label{equ:rate}
		R\left(t \right) = \left\{ \begin{array}{ll}
			n , &t_{n+1} < t  \le  {t_{n}},\\
			0, &{t_1} \le t.
		\end{array} \right.	
	\end{equation}
	where $ t_{R_{\max}+1}= T_{\mathrm{e}}$.
	When the adaptive modulation scheme adopts the transmission rate of $ n $, the corresponding transmission period is from $t_{n+1} $ to $t_{n} $.
	\subsection{Maximum Likelihood Detector and Sub-optimum Detector }	
	In this subsection, we introduce the maximum likelihood detector and the sub-optimum detector, which can demodulate the received signal under imperfect CSI to compute the instantaneous BEP of the mm-wave UAV link under wobbling.
	Moreover, a tight upper bound on the BEP of the sub-optimum detector is derived.
	
	The maximum likelihood detector under imperfect CSI was designed in \cite{JZhang2020TComm} for generalized polarization-space modulation.
	In this work, we simplify it by neglecting the spatial domain and polarization state of the detector for reducing computation complexity.
	Accordingly, the maximum likelihood detector detects the \textit{M}-ary PSK or \textit{M}-ary QAM signals under imperfect CSI as follows,
	\begin{equation}
		\label{equ:MLdetector}
		\hat k  = \mathop{\arg \min} \limits_{ k \in\left\lbrace 1,2, \cdots, M \right\rbrace  } \left\{ \ln \sigma^2_{ \mathrm{e},k}+ \frac{\left\| \mathbf{y}\left(t \right) - \sqrt{\gamma} \mathbf{h}\left( T_\mathrm{e} \right)C\left( T_\mathrm{e},t\right) s_{k} \right\|^2}{\sigma^2_{\mathrm{e},k }} \right\},
	\end{equation}
	where $ s_k $ is the $ k $-th constellation point in the \textit{M}-ary PSK or \textit{M}-ary QAM modulation, $ k \in \left\{1, 2, ..., M\right\} $,
	\begin{equation}
		\label{equ:sigma_effective}
		\sigma^2_{\mathrm{e},k }\!=\!  \gamma \left( 1 - \left[ C\left( T_\mathrm{e},t\right)\right] ^2 \right){\left| s_k\right|}^2 + 1,
	\end{equation}
	and $\sigma^2_{ \mathrm{e},k} $ is the effective covariance of the $ k $-th constellation point containing the power differences among symbols in a constellation diagram.
	
	The computational complexity of the maximum likelihood detector is still very high because of its searching process and complex computation process.  
	The computational complexity of the detector could be reduced further by neglecting the power differences among symbols in a constellation diagram \cite{JZhang2017TVT}, i.e., $ \sigma^2_{ \mathrm{e},k } =1$.
	Accordingly, the sub-optimum detector under imperfect CSI is given by
	\begin{equation}
	\label{equ:subOR}
	\hat m = \mathop {\arg \min }\limits_{ m \in \left\lbrace 1, 2, \cdots, M \right\rbrace  } \left\lbrace  {\left\| \mathbf{y}\left( t \right) - \sqrt{\gamma} \mathbf{h}\left( T_\mathrm{e} \right)C\left( T_\mathrm{e},t\right) {s_m}\right\| }^2 \right\rbrace ,
	\end{equation}
	where $ s_m $ is the $ m $-th constellation point in the \textit{M}-ary PSK or \textit{M}-ary QAM modulation, $ m \in \left\{1, 2, ..., M\right\} $.
		
	The UUB technique \cite{JZhang2020TComm} is employed to derive a tight upper bound on the BEP of the sub-optimum detector for the mm-wave UAV A2G link under imperfect CSI, which is provided in Theorem \ref{Th:ABEP}.

	\begin{theorem}\rm
		\label{Th:ABEP}
		The BEP of the sub-optimum detector for the mm-wave UAV A2G link under imperfect CSI is upper bounded by
		\begin{equation}
			\label{equ:ABEP_ana}
			\mathrm{BEP} \le \sum\limits_{m = 1}^M {\sum\limits_{\hat m = 1}^M {\frac{\mathcal{N}_{\left( 	m \to \hat m \right)}\mathrm{P}_{\left(m \to \hat m \right)}}{M{\log }_2\left( M \right)}}} \equiv \mathrm{UUB},
		\end{equation}
		where we define UUB, $ \mathcal{N}_{\left( {m \to \hat m} \right)} $ is the Hamming distance between symbols $ s_m$ and $s_{\widehat m} $, $ s_{\widehat{m}} $ is the modulation symbol determined by the sub-optimum detector, $ \mathrm{P}_{\left(m \to \hat m \right)} $ is the pairwise error probability (PEP), and is given by
		\begin{equation}
			\label{equ:PEP}
			\mathrm{P}_{ \left( m \to \hat{m} \right)} =Q\left( \sqrt { \frac{{\left\|\sqrt{\gamma} \mathbf{h}\left( T_\mathrm{e} \right)C\left( T_\mathrm{e},t\right)  \right\|}^2{\left| s_m - s_{\hat m} \right|}^2}{2{\gamma \left( 1 -C\left( T_\mathrm{e},t\right) ^2 \right){\left| s_m \right|}^2 + 2}}}\right).
		\end{equation}
	\end{theorem}
	
	\begin{IEEEproof}
		See Appendix A.
	\end{IEEEproof}
	
\begin{figure}[t]
		\centering
		\subfigure[PSK]{\includegraphics[width=0.45\linewidth]{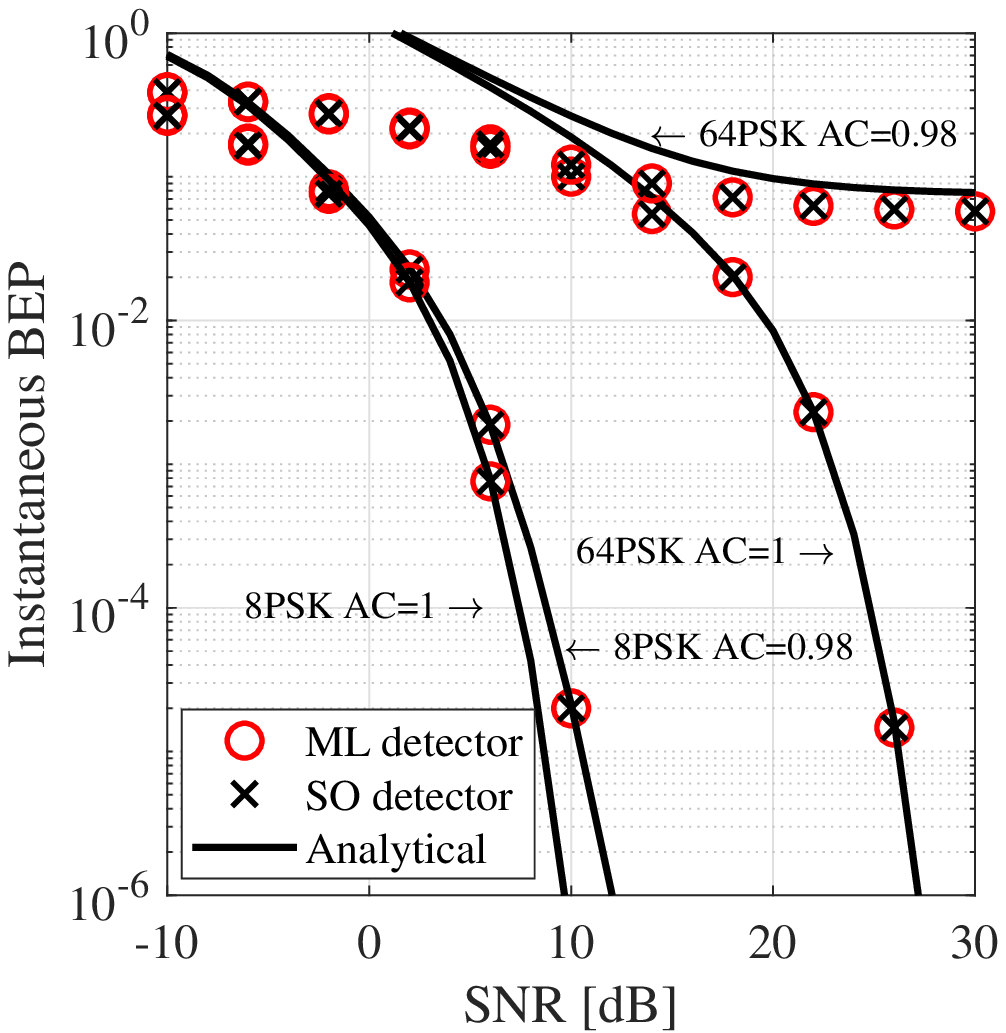}}
		\subfigure[QAM]{\includegraphics[width=0.45\linewidth]{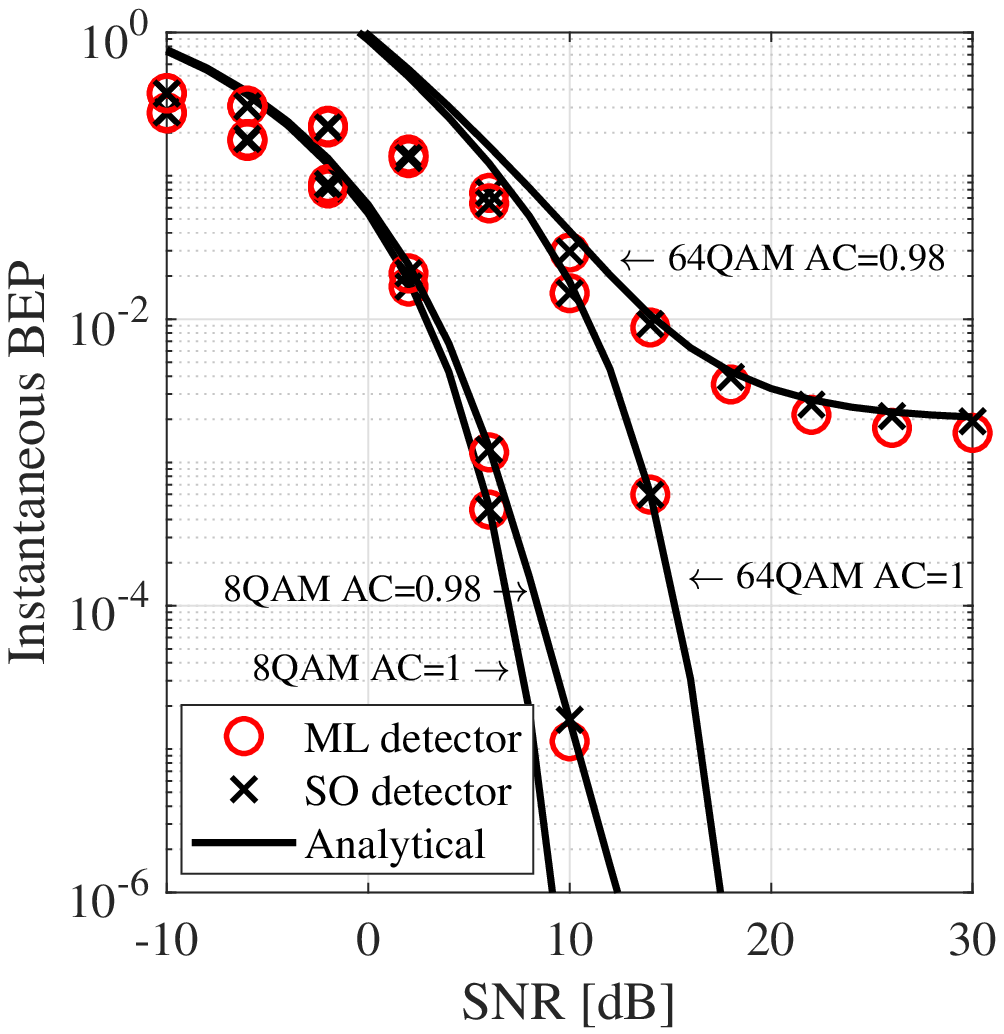}}
		\caption{The performance comparison of maximum likelihood and sub-optimum detector for PSK or QAM scheme. Markers show simulation results and solid lines illustrate analytical results. ML and SO denote maximum likelihood and sub-optimum, respectively. AC denotes the value of temporal autocorrelation.}
		\label{fig:PlanComparison}
	\end{figure}

	The performance of the maximum likelihood detector and the sub-optimum detector comparison with \textit{M}-ary PSK or \textit{M}-ary QAM signals are shown in Fig. \ref{fig:PlanComparison}. The analytical results computed by \eqref{equ:ABEP_ana} are compared with the Monte-Carlo simulations of the maximum likelihood detector and the sub-optimum detector for $ N_\mathrm{R} = 8 $, where \eqref{equ:MLdetector} and \eqref{equ:subOR} are used in the simulation, respectively. 
	The values of $ \mathbf{h}\left( T_{\mathrm{e}}\right) $ are listed in the Appendix B, where the value of case 1 is used for analytical and simulation analysis.
	
	The following observations are made therein.
	\begin{itemize}
		\item The BEP of the sub-optimum detector is close to that of the maximum likelihood detector at all SNR regimes. 
		Especially, when the \textit{M}-ary PSK signal is used, the performance of the maximum likelihood detector and the sub-optimum detector is the same because the modulus value of the \textit{M}-ary PSK signal is 1.
		\item Under both perfect and imperfect CSI, the analytical upper bound on the BEP of the sub-optimum detector is close to the BEP obtained from the simulations in the high SNR regime.
		Therefore, \eqref{equ:ABEP_ana} can be used to approximately compute the instantaneous BEP $ \beta_{i} $ of the adaptive modulation algorithm for the mm-wave UAV A2G link in the high SNR regime.
		\item 	The \textit{M}-ary QAM signals shows better BEP performance than that of the \textit{M}-ary PSK signals in the two employed detectors under both perfect and imperfect CSI. 
		\item The BEP will be significantly degraded by the imperfect CSI, especially at the high modulation order.
	\end{itemize}
	\section{Problem Formulation of Optimum Transmission Time and Its Solution}
	\subsection{Problem Formulation}
\begin{figure}[t]
	\centering
	\includegraphics[width=0.6\linewidth]{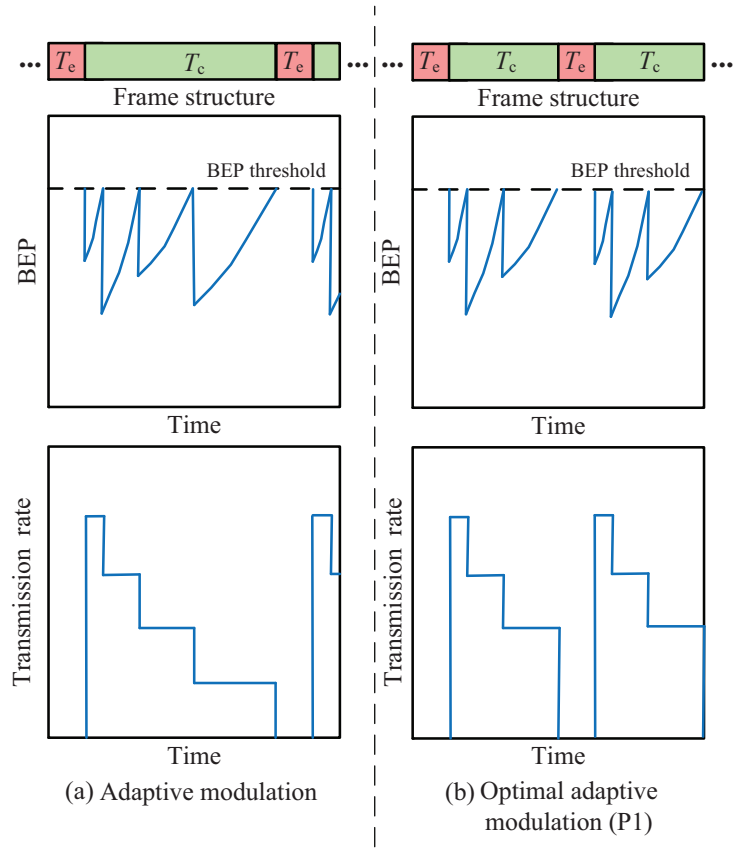}
	\caption{The schematic diagram of the problem (P1) and the potential solutions. }
	\label{fig:problemstatementad}
	\end{figure}

{\color{blue}	We formulate the following optimization problem to maximize the average transmission rate of the mm-wave UAV A2G link under UAV wobbling with a constant transmission power by optimizing the transmission time $ T_\mathrm{c} $ subject to the BEP threshold, i.e.,
	\begin{subequations}
		\begin{align}
			(\mathrm{P1}): \quad &  \mathop{\max}\limits_{ T_{\mathrm{c} }} \quad R_{\mathrm{ave}}\left( T_{\mathrm{c}} \right), \tag{12}\label{eq:OP_max}\\
			& \mathrm{s.t.} \quad   \beta_{i}  \le  \beta_{\mathrm{th}}, \label{eq:P1_th}\\
			& \enskip   \qquad 0 \le T_\mathrm{c} \le T_{\mathrm{co}}-T_\mathrm{e}, \label{eq:P1_tc}\\
			& \enskip \qquad P_\mathrm{T} = P_\mathrm{max}, \label{eq:P1_Pmax}
		\end{align}
	\end{subequations}
	where $T_{\mathrm{co}} $ denotes the channel coherence time; \eqref{eq:P1_th} requires that the instantaneous BEP $ \beta_{i} $ computed by \eqref{equ:ABEP_ana} must be kept below the predetermined BEP threshold  $ \beta_{\mathrm{th}}$; \eqref{eq:P1_tc} is the non-negative constraint on the transmission time; and \eqref{eq:P1_Pmax} requires that the transmission power is set at the maximum transmission power, which can support the maximum modulation order of the adaptive modulation under imperfect CSI. Fig. \ref{fig:problemstatementad} shows a schematic diagram of the optimization of $ T_\mathrm{c} $.}

	\subsection{Problem (P1) Solution: Optimum Adaptive Modulation Scheme}
	In this subsection, the algorithm to find the optimum transmission time as a solution to problem (P1) is presented.	
	\begin{algorithm}[t]
		\caption{The optimum transmission time finder.}
		\label{alg:max}
		\KwIn{$ T_{\mathrm{e}}$,  $ t_n $, $ R\left( t \right) $, $ t $. }
		\KwOut{$ R_\mathrm{ave,max} $, $ T_{\mathrm{max}} $.}
		$ R' \leftarrow$ \eqref{equ:der_Rave} with $ t $ and $ T_{\mathrm{e}}$\;
		\eIf{$ R' ==0 $}{
			$T_{\mathrm{max}} \leftarrow T_{\mathrm{c}}$\;}
		{
		$ T_{\mathrm{max}} $ $\leftarrow \mathrm{max}\left(\mathrm{find} \left(  R' < 0\right) \right) $\;	}
		$R_{\mathrm{ave, max}} \leftarrow$ \eqref{equ:Rave} with $ R(t) $, $T_{\mathrm{max}}$ and $ T_{\mathrm{e}}$\;
	\end{algorithm}
	
	{\color{blue} According to \eqref{eq:Ravedef} and \eqref{equ:rate}, the average transmission rate regarding $ T_\mathrm{c}= t - T_{\mathrm{e}} $ and the corresponding transmission time of adaptive modulation $ t_n $ can be written as 
	\begin{equation}
		\label{equ:Rave}
		R_{\mathrm{ave}}\left( T_{\mathrm{c}} \right) = \left\{ \begin{array}{ll}
			\frac{ t_{R_{\max}}+t_{R_{\max}-1}+ \cdots + t_{n+1}-R_{\max}T_{\mathrm{e}} }{T_{\mathrm{c}} + T_{\mathrm{e}}}+R(t) , &t_{n+1} \!<\! t\! \le \! {t_{n}},\\
			\frac{t_{R_{\max}}+ \cdots +t_1-R_{\max}T_{\mathrm{e}}}{T_{\mathrm{c}} + T_{\mathrm{e}}}, &{t_1} \!\le\! t.
		\end{array} \right.
	\end{equation}}
	
	We take the derivation of \eqref{equ:Rave} with respect to $T_{\mathrm{c}} $ and obtain
	\begin{equation}
		\label{equ:der_Rave}
		\frac{\mathrm{d} R_{\mathrm{ave}}\left(T_{\mathrm{c}}\right)}{\mathrm{d}T_{\mathrm{c}}} = \left\{ \begin{array}{ll}
			\frac{R_{\max}T_{\mathrm{e}} - t_{R_{\max}} -\cdots- t_{n+1}}{{\left( T_{\mathrm{c}} + T_{\mathrm{e}} \right)}^2}, &t_{n+1} \!<\! t\! \le \! {t_{n}},\\
			- \frac{t_{R_{\max}}+ \cdots +t_1-R_{\max}T_{\mathrm{e}}}{{\left( T_{\mathrm{c}} + T_{\mathrm{e}} \right)}^2}, &{t_1} \!\le\! t.
		\end{array} \right.
	\end{equation}
	
	\begin{figure}[t]
		\centering
		\includegraphics[width=0.45\linewidth]{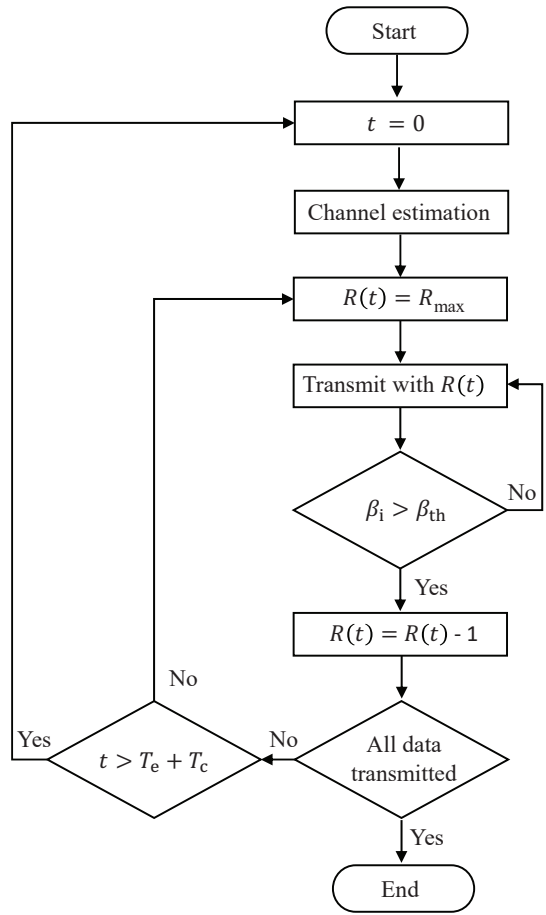}
		\caption{Flow chart of the optimum adaptive modulation scheme under imperfect CSI with time increase.}
		\label{fig: flow}
	\end{figure}
	The signal transmission time to maximize average transmission rate  $T_{\mathrm{max}}$  can be computed by the basic derivative property based on \eqref{equ:der_Rave}. 
	If the derivation of $R_{\mathrm{ave}}$ does not have a zero point, the first negative value of derivation will be detected as $T_{\mathrm{max}}$.
	When $T_{\mathrm{max}}$ is founded, the value of maximum average transmission rate $R_{\mathrm{ave,max}}$ can be computed by \eqref{equ:rate} and \eqref{equ:Rave}, respectively.
	The pseudo-code of the proposed optimum adaptive modulation scheme is given in Algorithm  \ref{alg:max}, where $ R' $ is a local variable.
	
	The flow chart of the optimum adaptive modulation scheme of the mm-wave UAV A2G link under imperfect CSI  is shown in Fig. \ref{fig: flow}.
		Finally, the problem (P1) is solved by optimizing the transmission time followed the adaptive modulation scheme and Algorithm \ref{alg:max}, where the ended transmission rate of optimum adaptive modulation $ R_\mathrm{op}$ and the maximum average transmission rate $ R_\mathrm{ave,max}$ can be computed.

	\section{Problem Formulation of Power Control Policy and Its Solution}
	\subsection{Problem Formulation}
\begin{figure}[t]
		\centering
		\includegraphics[width=0.6\linewidth]{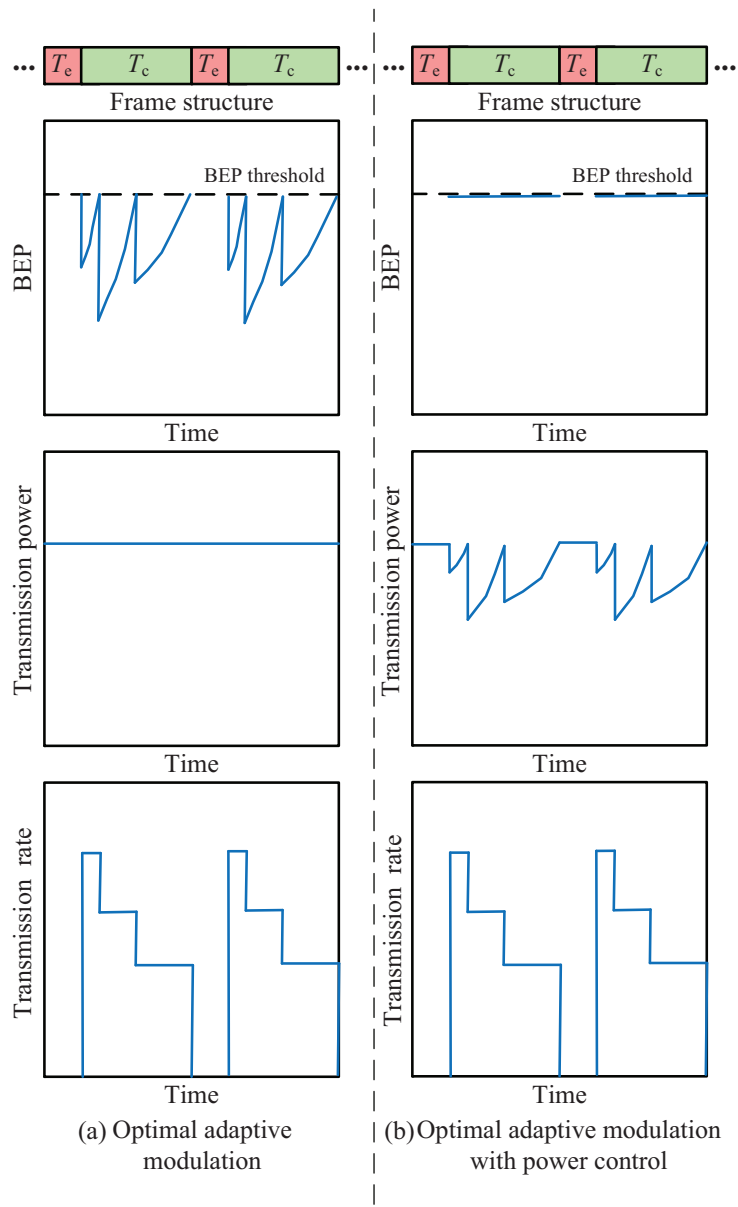}
		\caption{The schematic diagram of the problem (P2) and the potential solutions. }
		\label{fig:problemstatementpc}
	\end{figure}

	We formulate the following optimization problem to minimize the transmission power of the mm-wave UAV A2G link under UAV wobbling maintaining maximized average transmission rate subject to the BEP threshold, i.e.,
	\begin{subequations}
		\begin{align}
			(\mathrm{P2}): \quad & \mathrm{min} \quad P_\mathrm{T} (t), \tag{17}\\	
			\quad & \mathrm{s.t.} \quad 0 \le \ t \le T_\mathrm{e}+T_\mathrm{c} , \label{eq:P2_time}\\
			& \enskip \qquad R_{\mathrm{ave} }=  R_\mathrm{ave,max}, \label{eq:P2_rate}\\
			& \enskip \qquad \beta_{i} \le  \beta_{\mathrm{th}}, \label{eq:P2_th}\\
			& \enskip \qquad 0 < P_\mathrm{T}( t ) \le P_\mathrm{max}, \label{eq:P2_power}
		\end{align}
	\end{subequations}
	where \eqref{eq:P2_time} is the constraint on power control policy adopting period; \eqref{eq:P2_rate} requires the average transmission rate is set at the maximized average transmission rate following optimum adaptive modulation; \eqref{eq:P2_th} requires that the instantaneous BEP $ \beta_{i} $ of the wobbling mm-wave UAV A2G link must be kept below the predetermined BEP threshold  $ \beta_{\mathrm{th}}$; \eqref{eq:P2_power} is the adaptive transmission power constraint on maximum transmission power.
	Fig. \ref{fig:problemstatementpc} shows a schematic diagram of the transmission power optimization.
	\subsection{Problem (P2) Solution: Power Control Policy}
	In this subsection, we design the power control policy with varying transmission power to solve the problem (P2). 
	 
	The SNR is a key parameter to bridge the transmission power and the system performance. 
	The minimum instantaneous transmission power can be achieved when the minimum required instantaneous SNR is employed in the transmission frame. 
	Although it is hard to get the value of the instantaneous SNR directly from \eqref{equ:ABEP_ana}, an alternative expression of BEP exists to reduce the complexity of computation.
	According to \cite{Lu1999TComm}, the signal-space concept could be used to obtain accurate BEP approximations of \eqref{equ:ABEP_ana} for \textit{M}-ary PSK modulation scheme, where the Gray code mapping property describes that an error in an adjacent symbol is accompanied by one and only one bit error.
	For \textit{M}-ary PSK and $ |s_m|=1 $, \eqref{equ:ABEP_ana} based on signal-space concept \cite{Lu1999TComm} can be rewritten as 
	\begin{equation}\label{equ:PSK_app}
	\beta_{\mathrm{th}} =
	\left\{ \begin{array}{ll}
	\mathrm{Q}\left( \sqrt { \frac{2{\left\|\sqrt{\gamma} \mathbf{h}\left( T_\mathrm{e} \right)C\left( T_\mathrm{e},t\right)  \right\|}^2 }{{\gamma \left( 1 -C\left( T_\mathrm{e},t\right) ^2 \right){\left| s_m \right|}^2 + 1}}}\right),& M=2,\\
	\frac{2}{\log_2(M)}\mathrm{Q}\left( \sqrt { \frac{{\left\|\sqrt{\gamma} \mathbf{h}\left( T_\mathrm{e} \right)C\left( T_\mathrm{e},t\right)  \right\|}^2{\left(1-\cos\left( \frac{2\pi}{M} \right)  \right) }}{{\gamma \left( 1 -C\left( T_\mathrm{e},t\right) ^2 \right){\left| s_m \right|}^2 + 1}}}\right),&  M >2.
		\end{array}	\right.
	\end{equation}

	\begin{figure}[t]
	\centering
	\includegraphics[width=0.45\linewidth]{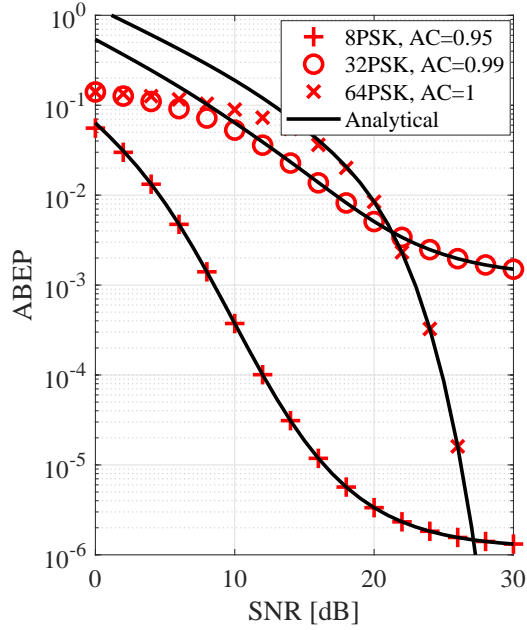}
	\caption{The BEP approximation of \eqref{equ:ABEP_ana} employing \textit{M}-ary PSK. Markers show approximate results and solid lines illustrate analytical results. AC denotes the value of temporal autocorrelation.}
	\label{fig:ABEP_approx}
	\end{figure}

	Fig. \ref{fig:ABEP_approx} shows the BEP approximations of \textit{M}-ary PSK modulation scheme are close to the analytical upper bound on the BEP of the sub-optimum detector in high SNR regime.
	The analytical and approximation results are computed by \eqref{equ:ABEP_ana} and \eqref{equ:PSK_app}, respectively. 

	We take the derivation of \eqref{equ:PSK_app} with respect to the BEP threshold $ \beta_{\mathrm{th}} $ and obtain the instantaneous SNR, 
	\begin{equation}\label{equ:PSK_gamma}
		\gamma_{\min} =
		\left\{ \begin{array}{ll}
			\frac{{\mathrm{Q}^{-1}\left( \beta_{\mathrm{th}}\right)}^2 }{2{\left\|\mathbf{h}\left( T_\mathrm{e} \right)C\left( T_\mathrm{e},t\right)  \right\|}^2-\left( 1 -C\left( T_\mathrm{e},t\right)^2\right){\mathrm{Q}^{-1}\left( \beta_{\mathrm{th}} \right)}^2},&M=2,\\
			\frac{\alpha^2}{{\left\|\mathbf{h}\left( T_\mathrm{e} \right)C\left( T_\mathrm{e},t\right)  \right\|}^2\left( 1-\cos(2\pi /M)\right) -\left( 1 -C\left( T_\mathrm{e},t\right)^2\right) \alpha^2}, &M>2,
		\end{array}
		\right.
	\end{equation}
	where $ \alpha=\mathrm{Q}^{-1}\left( \frac{\beta_{\mathrm{th}}\log_2(M)}{2}\right) $ and $ \mathrm{Q}^{-1}(\cdot) $ stands for the inverse function of the Q-function.

	For \textit{M}-ary QAM ($ M>2 $), the instantaneous SNR of the mm-wave UAV A2G link is the root of
	\begin{equation}
		\label{equ:QAM_app}
		\beta_{\mathrm{th}} = \sum\limits_{m = 1}^M {\sum\limits_{\hat m = 1}^M {\frac{\mathcal{N}_{\left( 	m \to \hat m \right)}\mathrm{P}_{\left(m \to \hat m \right)}}{M{\log }_2\left( M \right)}}}. 
	\end{equation}
	Then, the Newton-Raphson method \cite{Hildebrand1973} could be employed to derive the approximation root, which is provided in Lemma 1.	
	More specifically, Lemma 1 should be replicated multiple times to get an accurate approximation root of \eqref{equ:QAM_app} as the instantaneous SNR of the mm-wave UAV A2G link under wobbling.
	
	\begin{lemma}\rm
		\label{le:QAM_SNR}
		The approximation root of \eqref{equ:QAM_app} using the Newton-Raphson method is given by
				\begin{equation}\label{equ:QAM_SNR}
			\gamma_\mathrm{re+1}=\gamma_\mathrm{re}\left( \frac{u_m}{\beta_{\mathrm{th}}}\right)^{\frac{u_m}{v_m}}, 
		\end{equation}
	where
	\begin{equation}\label{equ:um}
		u_m=	\sum_{m=1}^{M}{	\sum_{\hat{m}=1}^{M}{	\frac{\mathcal{N}_{\left( 	m \to \hat m \right)} \mathrm{Q}\left( \sqrt{\frac{{\Lambda}\gamma_\mathrm{re}}{\psi \gamma_\mathrm{re}+2}} \right) }{M\log_2(M)}}},
	\end{equation}	
	\begin{equation}\label{equ:vm}
		v_m=\sum_{m=1}^M{\sum_{\hat{m}=1}^M{\frac{\mathcal{N}_{\left( 	m \to \hat m \right)} \Lambda\sqrt{\gamma_\mathrm{re}}\exp{\left( -\frac{\Lambda \gamma_\mathrm{re}}{2\psi \gamma_\mathrm{re}+4}\right) }}{M\log_2(M)\sqrt{2\pi \Lambda\left( \gamma_\mathrm{re}\psi+2\right)^3}} }	},
	\end{equation}
	\begin{equation}\label{equ:lamba}
	\Lambda=\left\|\mathbf{h}\left( T_\mathrm{e} \right)C\left( T_\mathrm{e},t\right)  \right\|^2 {\left| s_m - s_{\hat m} \right|}^2,
	\end{equation}
	\begin{equation}\label{equ:}
	\psi=2\left( 1 -C\left( T_\mathrm{e},t\right)^2\right) |s_m|^2,
	\end{equation}
	$ \gamma_\mathrm{re} $ stands the initial guess root, and $ \gamma_\mathrm{re+1} $ is the result of approximation root.
	\end{lemma}
	\begin{IEEEproof}
		See Appendix C.
	\end{IEEEproof}

	In general, the thermal noise power is calculated by $ N_0 \left[ \mathrm{dBm}\right]  = 10 \log_{10}(k_\mathrm{b}TB)$, where $ k_\mathrm{b}$ is the Boltzmann constant ($ 1.38\times 10^{-23}$ J/K), $T $ is the temperature in K, and  $ B $ is the bandwidth.
	Therefore, the minimum instantaneous transmission power is derived as 
	\begin{equation}\label{eq:pwraa}
		P_\mathrm{min}\left( t\right)  \left[ \mathrm{dBm}\right]  = {\gamma}_\mathrm{min}\left( t\right) +P_\mathrm{L}+N_0.
	\end{equation}

	{\color{blue} To support the adaptive modulation with the power control policy, the extra control message is required for the power control policy in the feedback information from the ground node to the UAV.
	The feedback information of adaptive modulation contains the initial transmission rate $ R_{\max} $ (one integral number) and the corresponding switching transmission time $ t_n $ (multiple floating numbers).
	Moreover, the extra feedback information of adaptive modulation with power control is the transmission power $ P(t) $ (multiple floating numbers).
	Accordingly, the size of feedback information of adaptive modulation with power control policy is nearly double of that without power control policy.}
	\section{Numerical Results}
	\begin{table}[t]
	\centering
	\caption{Simulation Parameters \cite{Zeng2019PIEEE, Agiwal2016TUT}}
	\begin{tabular}{|l|l|}
		\hline
		Horizontal distance ($\left\|\mathbf{w}\right\|$)& 100 m \\
		\hline
		Carrier frequency ($ f $) & 28 GHz\\
		\hline
		Number of antenna at receiver ($N_\mathrm{R} $)& 8\\
		\hline
		UAV height ($H$) & 100 m\\
		\hline
		BEP threshold ($ \beta_{\mathrm{th}} $)& $ 10^{-5} $ \\
		\hline
		 Channel estimation period ($ T_{\mathrm{e}} $)& $ 10^{-3} $ s\\
		 \hline
		 Maximum transmission power ($ P_\mathrm{max} $)& 35 dBm\\
		\hline
		Bandwidth ($ B $) & 100 MHz\\
		\hline
		Temperature ($T $)& 300 K\\
		\hline
	\end{tabular}
	\label{table:1}
	\end{table}

	In this section, the performance of the proposed optimum adaptive modulation scheme in conjunction with the power control policy is evaluated. 
	How the SNR and the BEP threshold impact the maximum transmission rate will be analyzed below. 
	How effective the power control policy of the optimum adaptive modulation scheme will be shown.
	
	\begin{figure}[t]
		\centering
		\subfigure[PSK]{\includegraphics[width=0.45\linewidth]{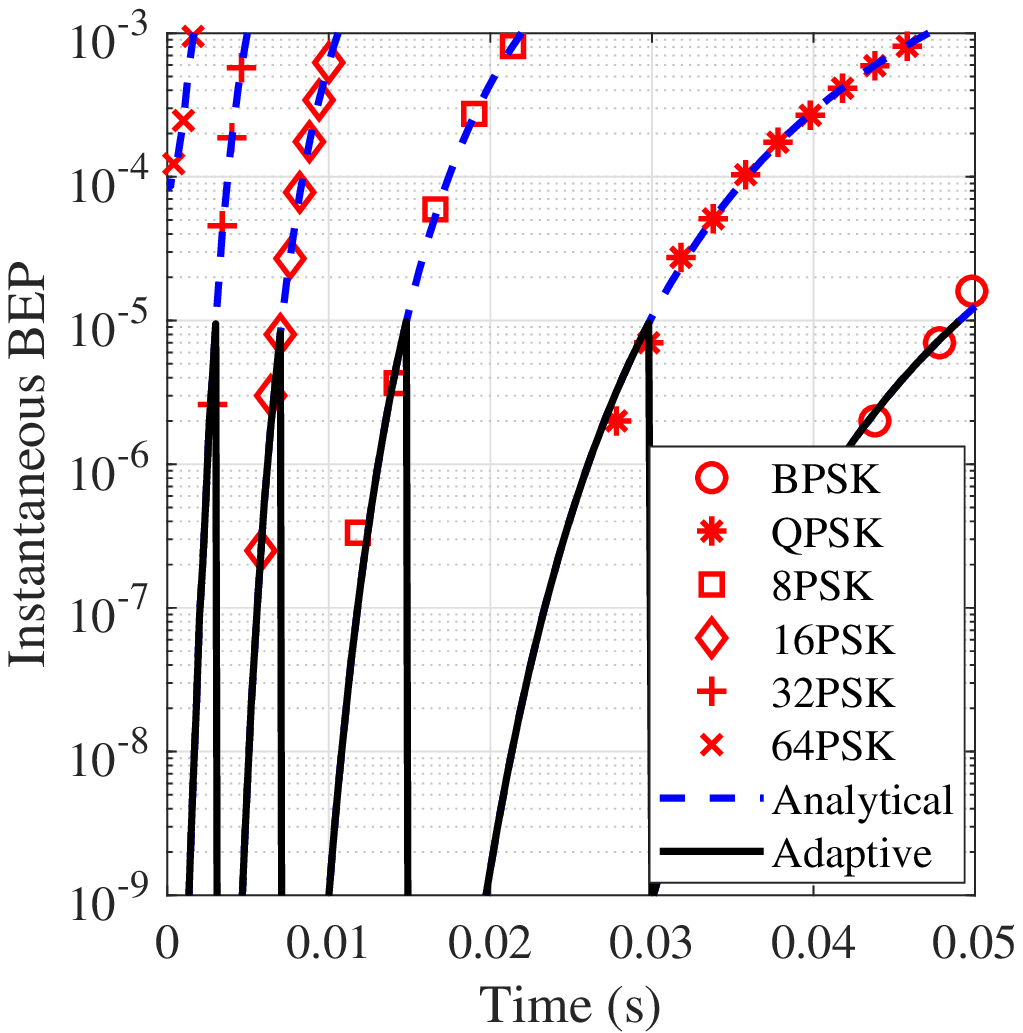}}
		\subfigure[QAM]{\includegraphics[width=0.45\linewidth]{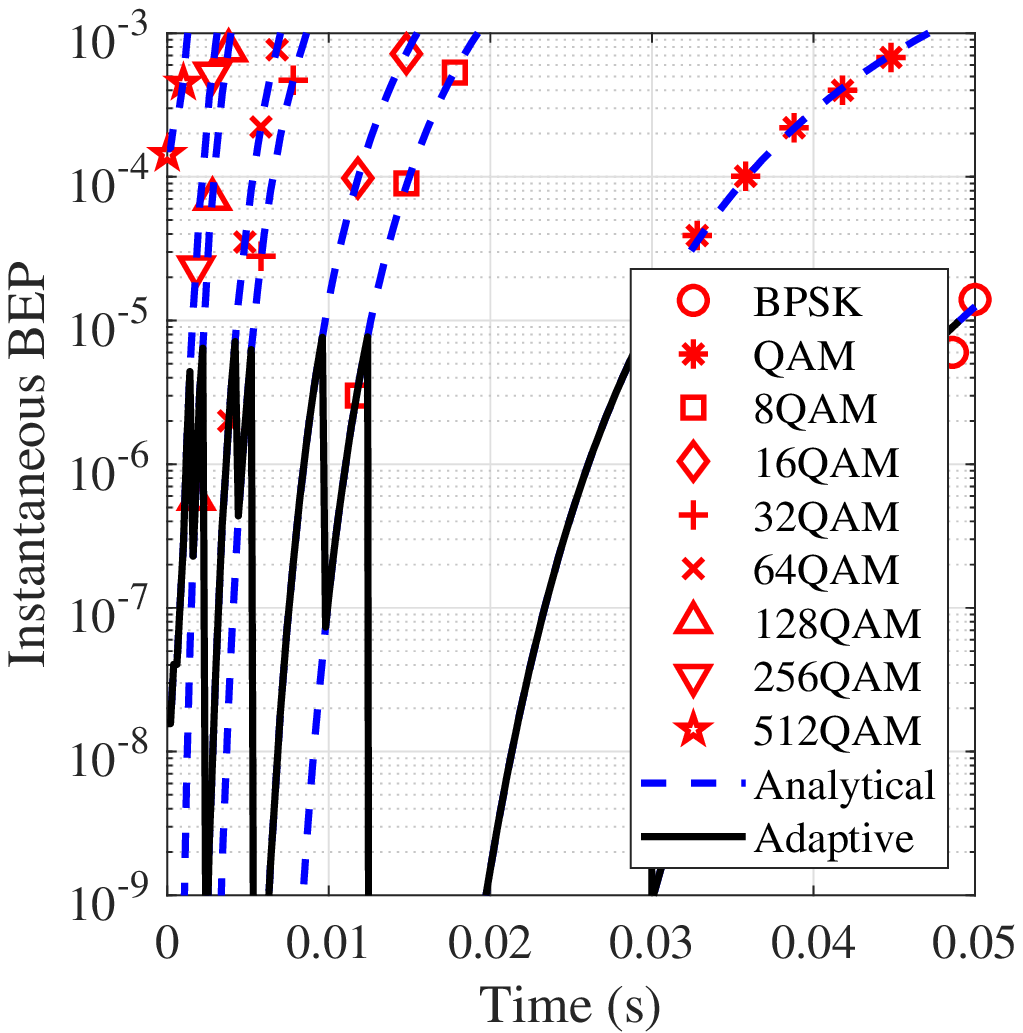}}
		\caption{Adaptive modulation scheme employing \textit{M}-ary PSK or \textit{M}-ary QAM under case 1. Markers show simulation results and dash lines illustrate analytical results. }
		\label{fig:MISO_Time_ABEP_1E5}
		\centering
		\subfigure[PSK]{\includegraphics[width=0.45\linewidth]{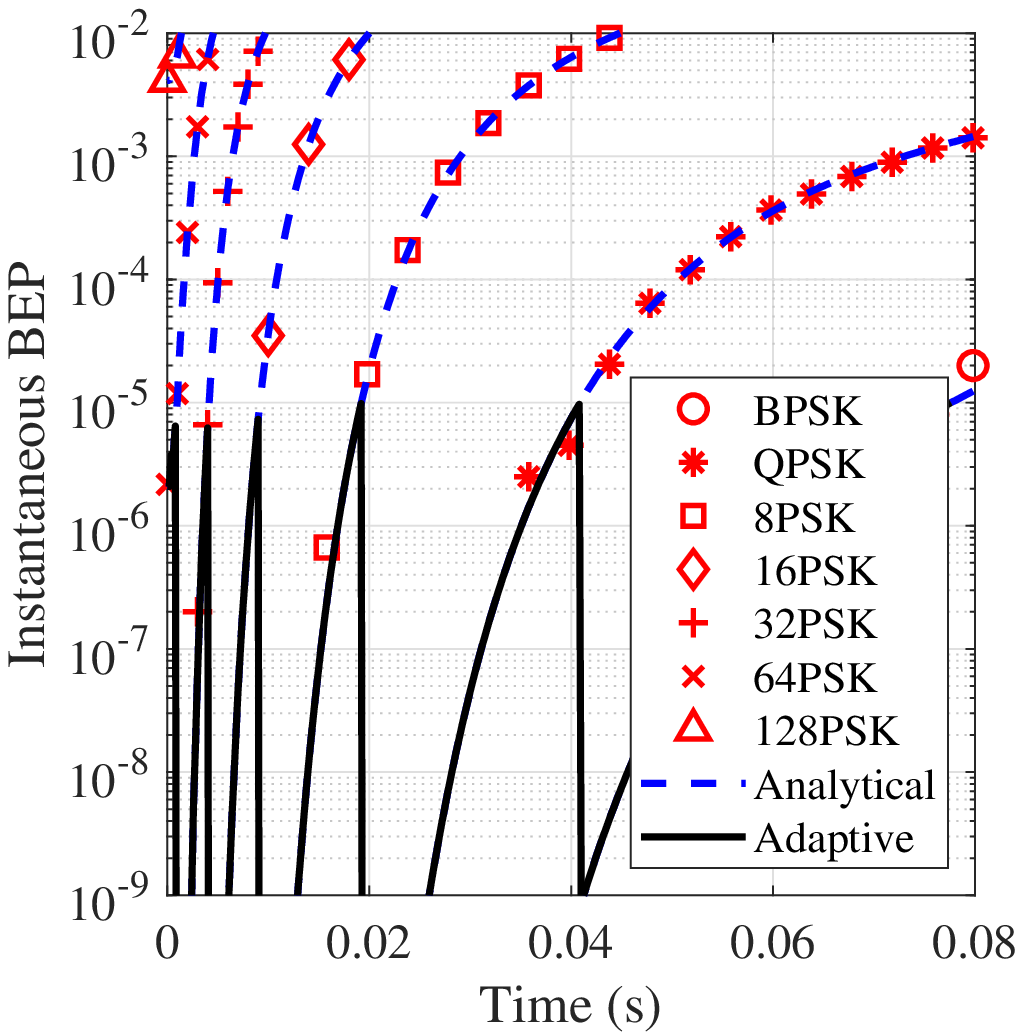}}
		\subfigure[QAM]{\includegraphics[width=0.45\linewidth]{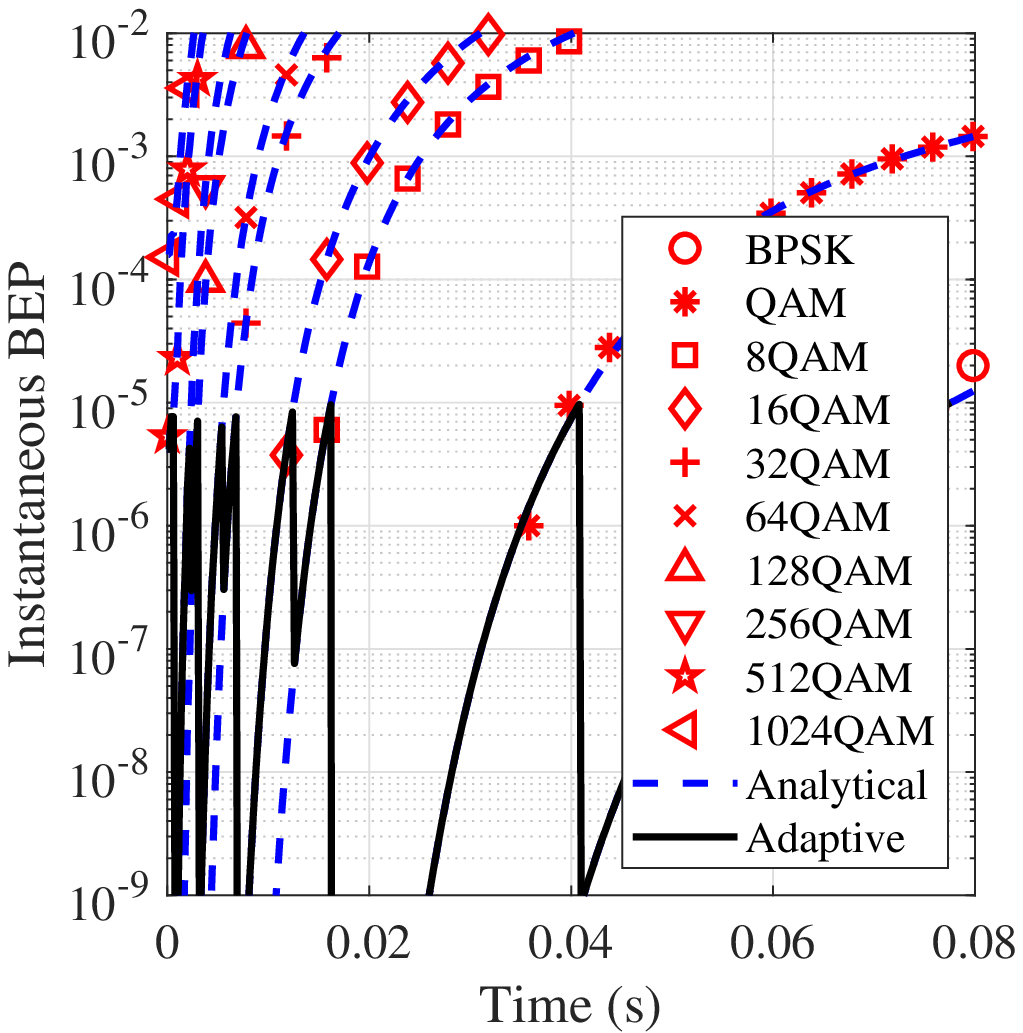}}
		\caption{Adaptive modulation scheme employing \textit{M}-ary PSK or \textit{M}-ary QAM under case 2. Markers show simulation results and dash lines illustrate analytical results. }
		\label{fig:MISO_Time_ABEP_1E52}
	\end{figure}
	
	We performed the simulations based on the temporal ACF of the mm-wave UAV A2G link is given by \cite{YangArxiv},
	\begin{equation}
		\label{equ:bessel}
		\begin{array}{l}
			C\left( \Delta t\right) = {e^{ - 0.5\sigma_{\mathrm{v}}^2{\left( \frac{\omega _{\mathrm{c}}}{c\left( \omega _{\mathrm{v}}^2 + {\mu ^2} \right)} \right)}^2 \left( \mu \Delta t\left( \omega _{\mathrm{v}}^2 + \mu ^2 \right) - 2\mu \omega _{\mathrm{v}}\sin \left( \omega _{\mathrm{v}}\Delta t \right) e^{ - \mu \Delta t} +\left( \mu ^2 -\omega _{\mathrm{v}}^2\right) \cos \left( \omega _{\mathrm{v}}\Delta t\right)e^{ - \mu \Delta t} - \mu ^2 + \omega _{\mathrm{v}}^2\right)  }}\\
			\hspace{0.45in}\times J_{0}\left( {j0.5\sigma_{\mathrm{v}}^2{\left( \frac{\omega _{\mathrm{c}}}{c} \right)^2}\frac{ \mu \sin \left( \omega _{\mathrm{v}}\Delta t \right) - \omega _{\mathrm{v}}\cos \left( \omega _{\mathrm{v}}\Delta t \right) + \omega _{\mathrm{v}}e^{  - \mu \Delta t } }{\left( \omega _{\mathrm{v}}^2 + {\mu ^2} \right)\omega _{\mathrm{v}}}} \right),
		\end{array}
	\end{equation}
	where $\Delta t = t-T_\mathrm{e}$, $J_{0}(\cdot)$ denotes the Bessel function of the first kind with an order zero, $\sigma^2_{\mathrm{v}} $ is the variance of the UAV movement velocity, $ \omega_{\mathrm{c}} $ is the carrier frequency, $ \omega_{\mathrm{v}} $ is the mechanical vibration frequency of the UAV (rad/sec), $ \mu $ is the parameter to measure how fast the envelope of velocity changes with time, and $ \sigma_{\mathrm{v}}^2 = (0.005)^2\frac{\omega_v^2+\mu^2}{\mu}$. 
	The typical scenarios setting for $ \omega_{\mathrm{v}} $  is $ 20\pi $ rad/sec  and $ \mu $ is 30, where the detailed explanation for \eqref{equ:bessel} setting could be found in \cite{YangArxiv}. 
	According to  \cite{Agiwal2016TUT} and \cite{Blogh2002}, the suggested values of the maximum tolerable BEP for the speech, video, and data signals ares $ 10^{-3}$, $10^{-5} $, and  $ 10^{-6}$, respectively.
	The value of estimated channel impulse response is shown in Appendix B. 
	The system parameters used in simulation are shown in Table \ref{table:1}. 
			
	\begin{figure}[t]
		\centering
		\subfigure[PSK]{\includegraphics[width=0.45\linewidth]{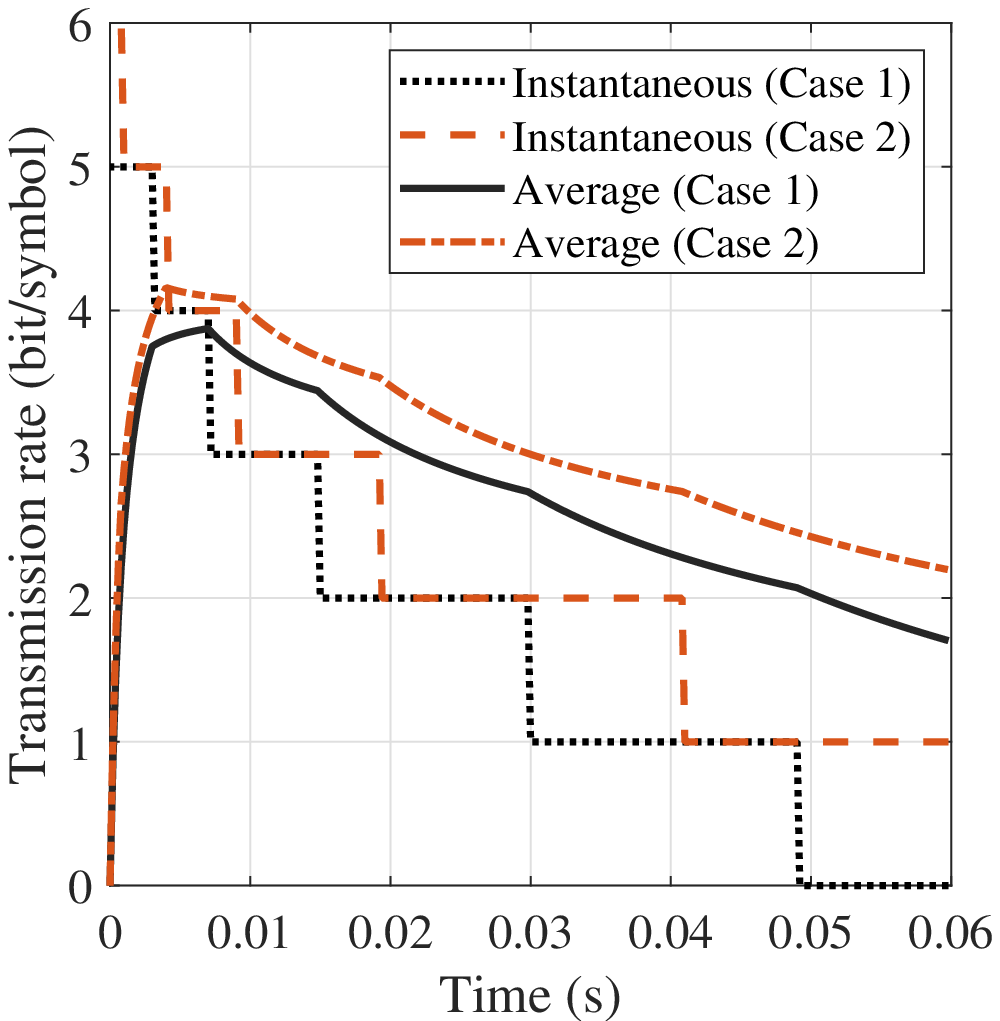}}
		\subfigure[QAM]{\includegraphics[width=0.45\linewidth]{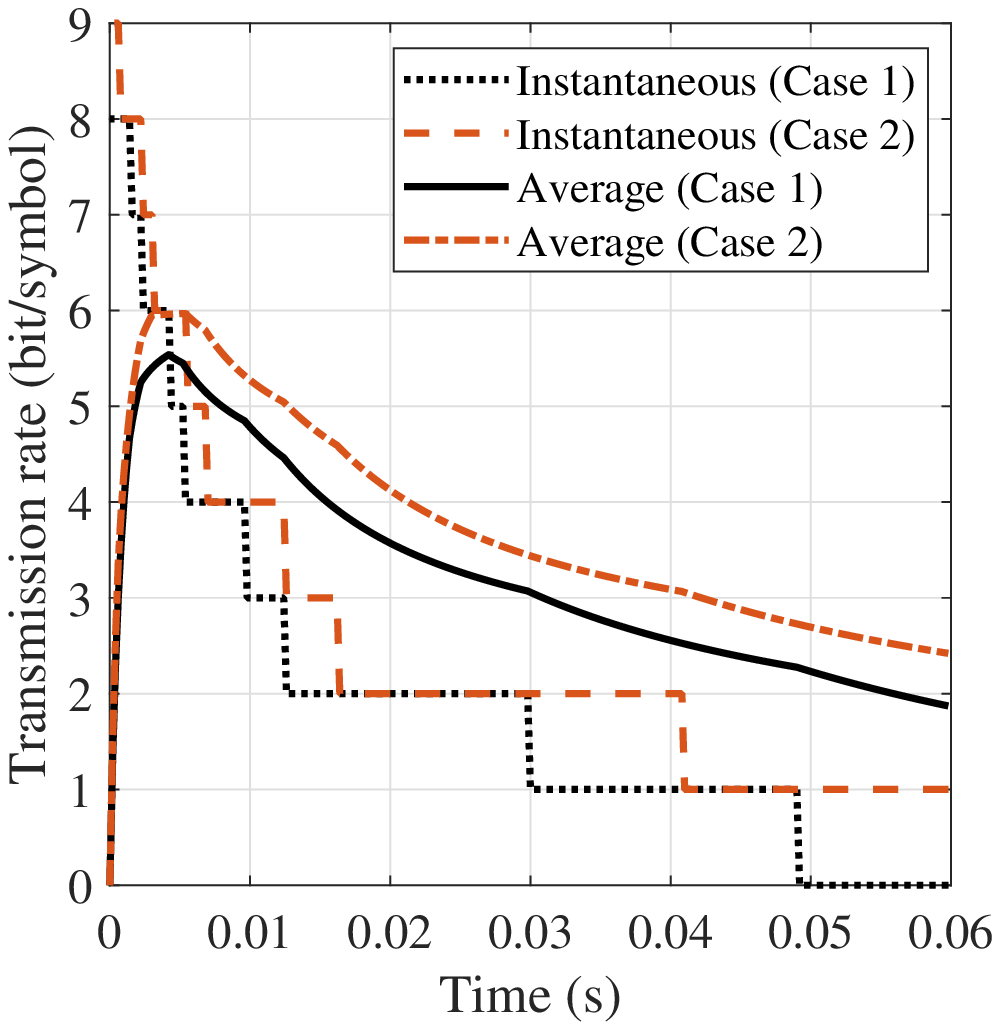}}
		\caption{The instantaneous and average transmission rate for two cases of $ \mathbf{h}(T_\mathrm{e}) $ with adaptive modulation scheme at BEP threshold $10^{-5} $.  }
		\label{fig:RateComparison}
	\end{figure}
	\begin{figure}[t]
		\centering
		\subfigure[PSK]{\includegraphics[width=0.45\linewidth]{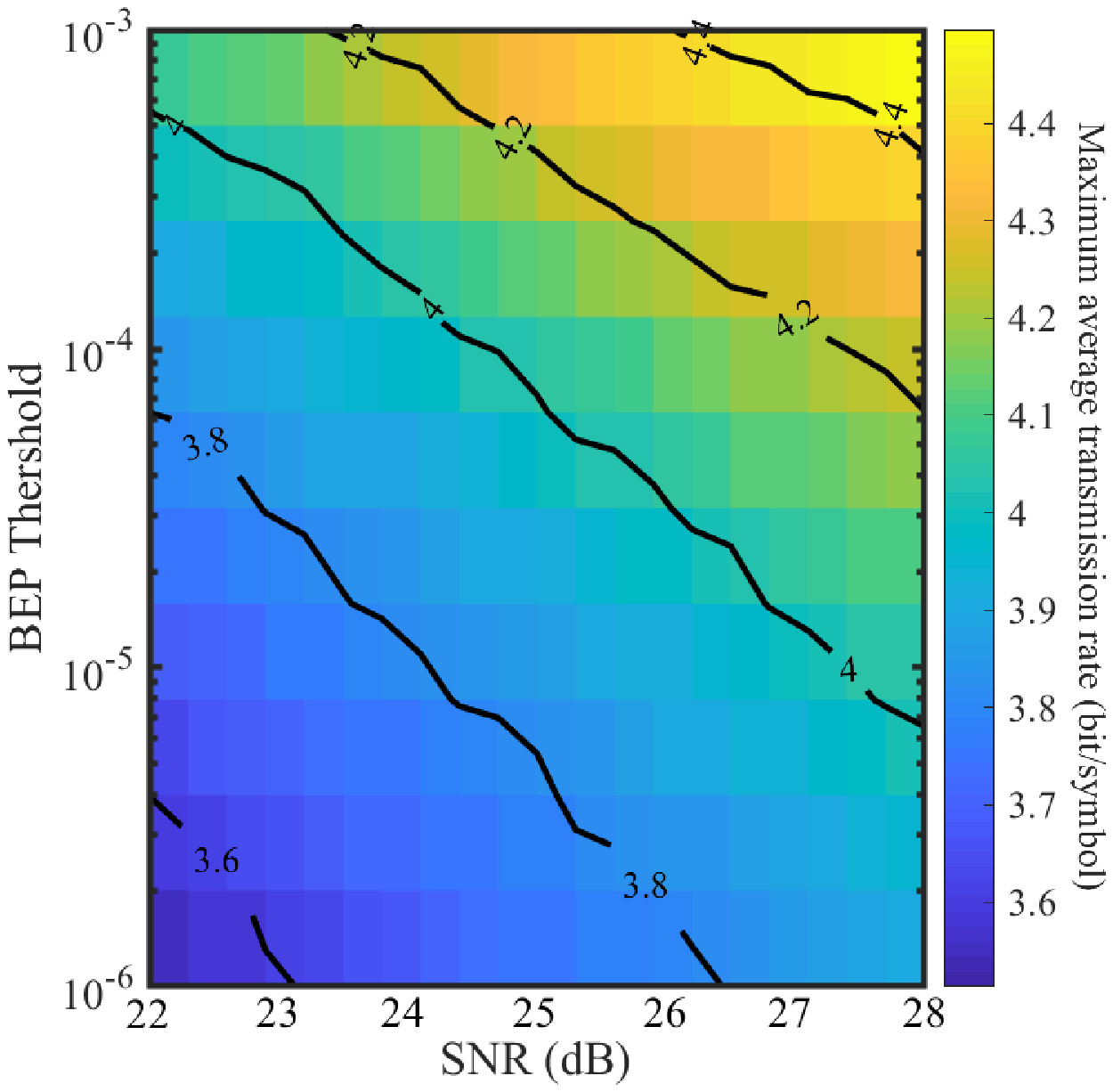}}
		\subfigure[QAM]{\includegraphics[width=0.45\linewidth]{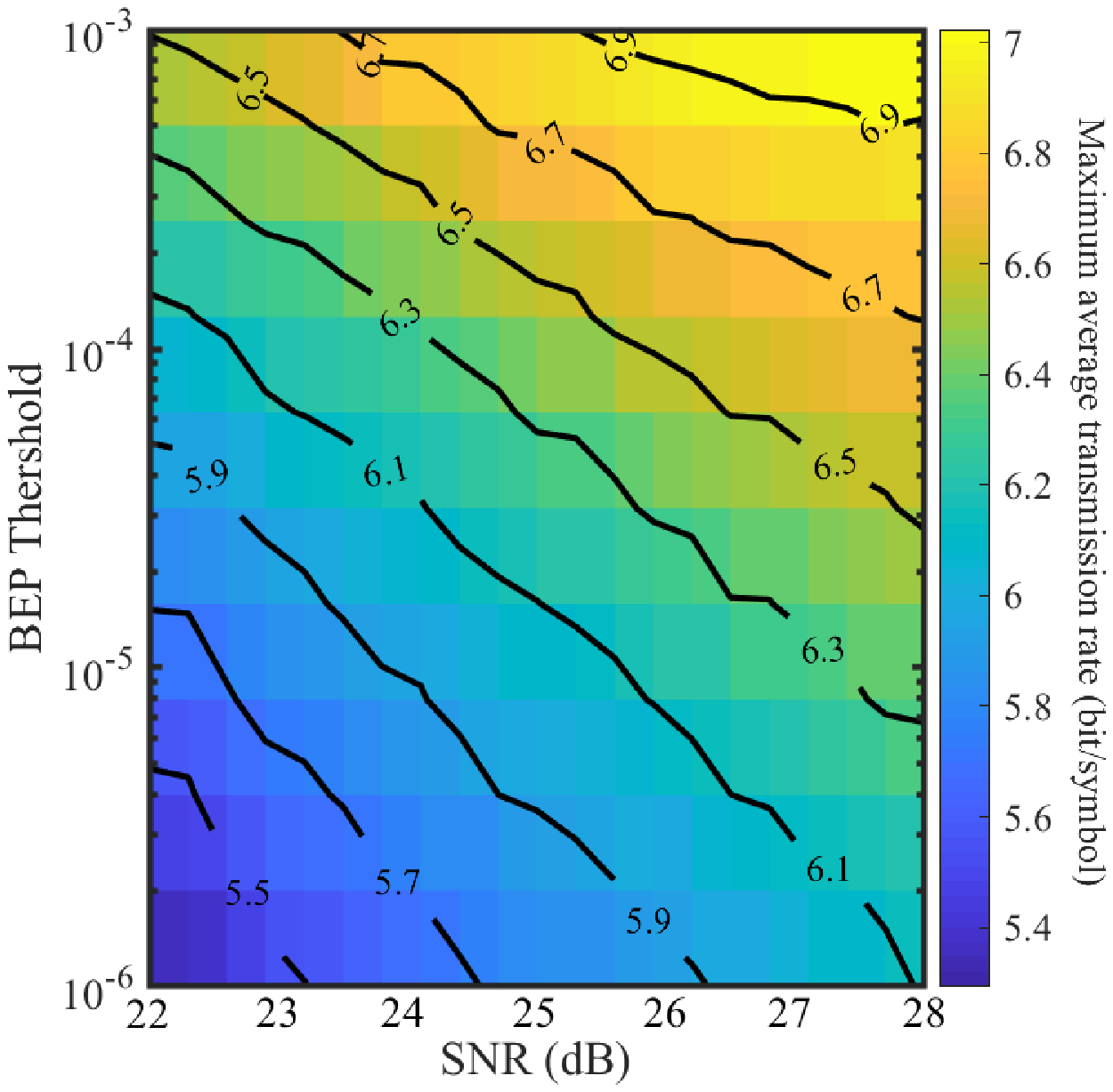}}
		\caption{Comparison of maximum average data rates for the designed adaptive modulation scheme employing \textit{M}-ary PSK or \textit{M}-ary QAM under different SNR and BEP thresholds. }
		\label{fig:Contour}
	\end{figure}
	
	\begin{figure}[ht]
		\centering
		\subfigure[Transmission power]{\includegraphics[width=0.45\linewidth]{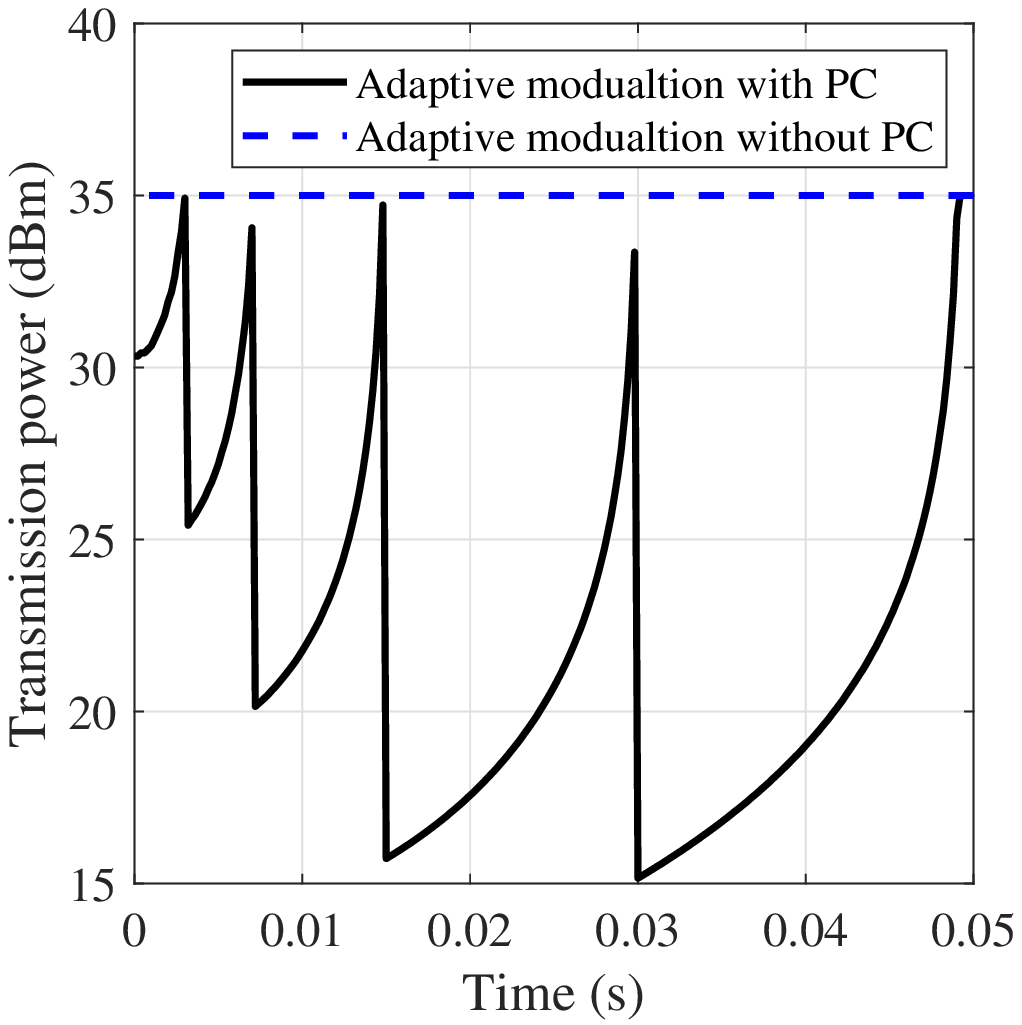}}
		\subfigure[BEP]{\includegraphics[width=0.45\linewidth]{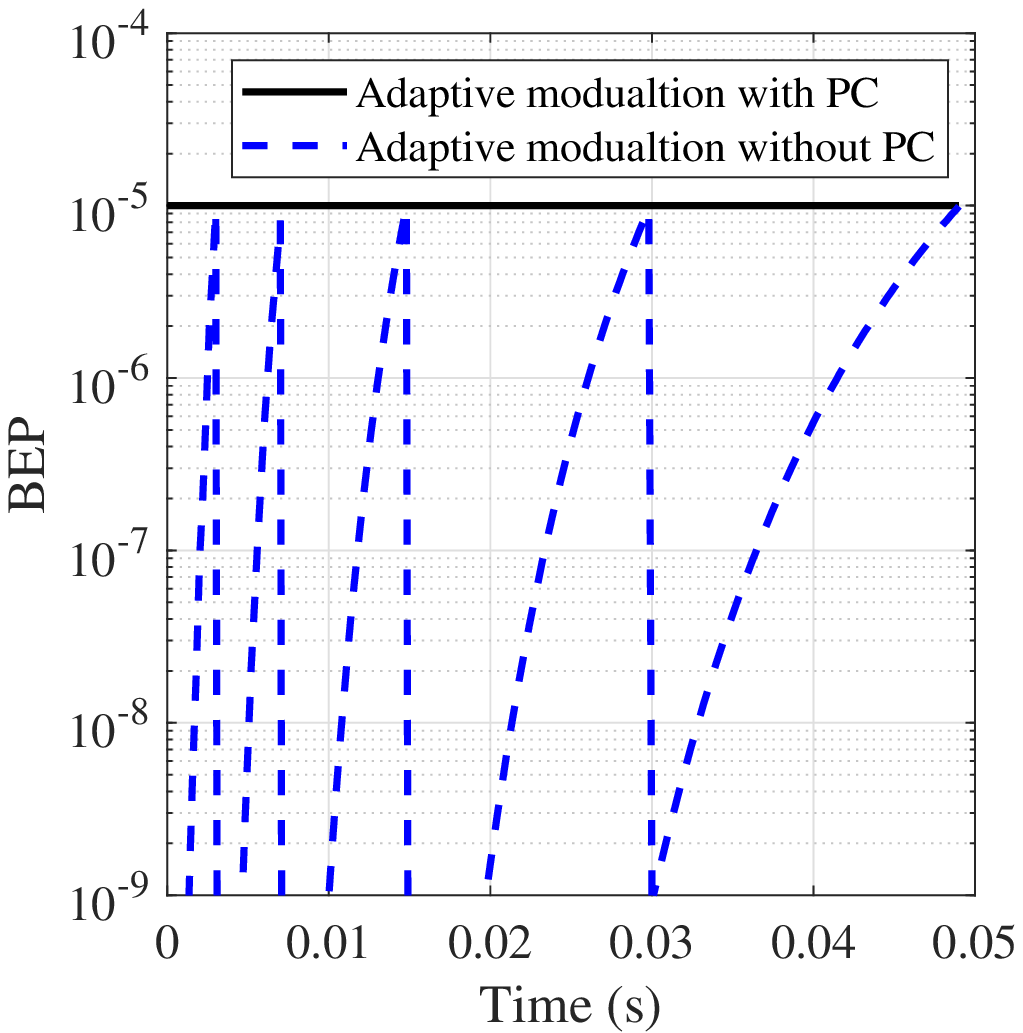}}
		\caption{The performance of power control policy for the adaptive modulation scheme for \textit{M}-ary PSK and BEP threshold at $ 10^{-5} $ under case 1.}
		\label{fig:PCPSK}
		\centering
		\subfigure[Transmission pwoer]{\includegraphics[width=0.45\linewidth]{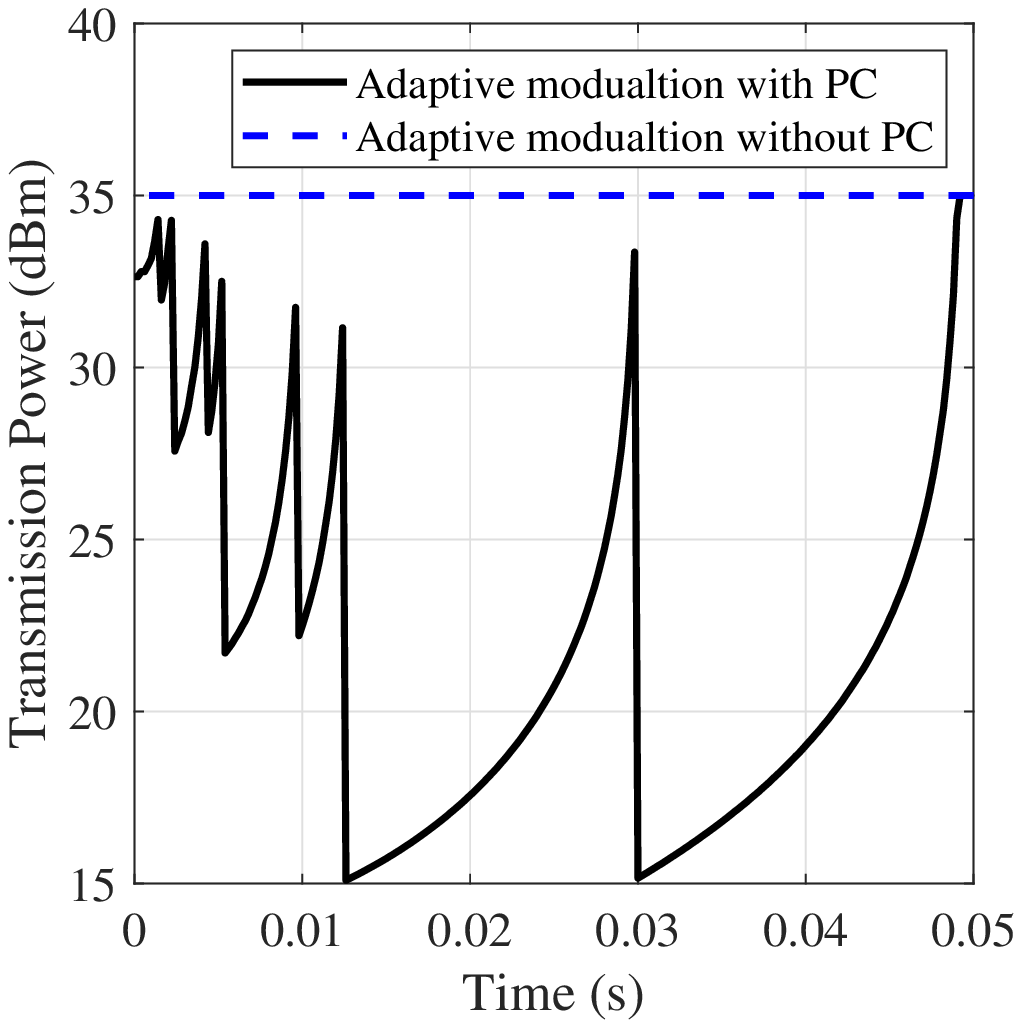}}
		\subfigure[BEP]{\includegraphics[width=0.45\linewidth]{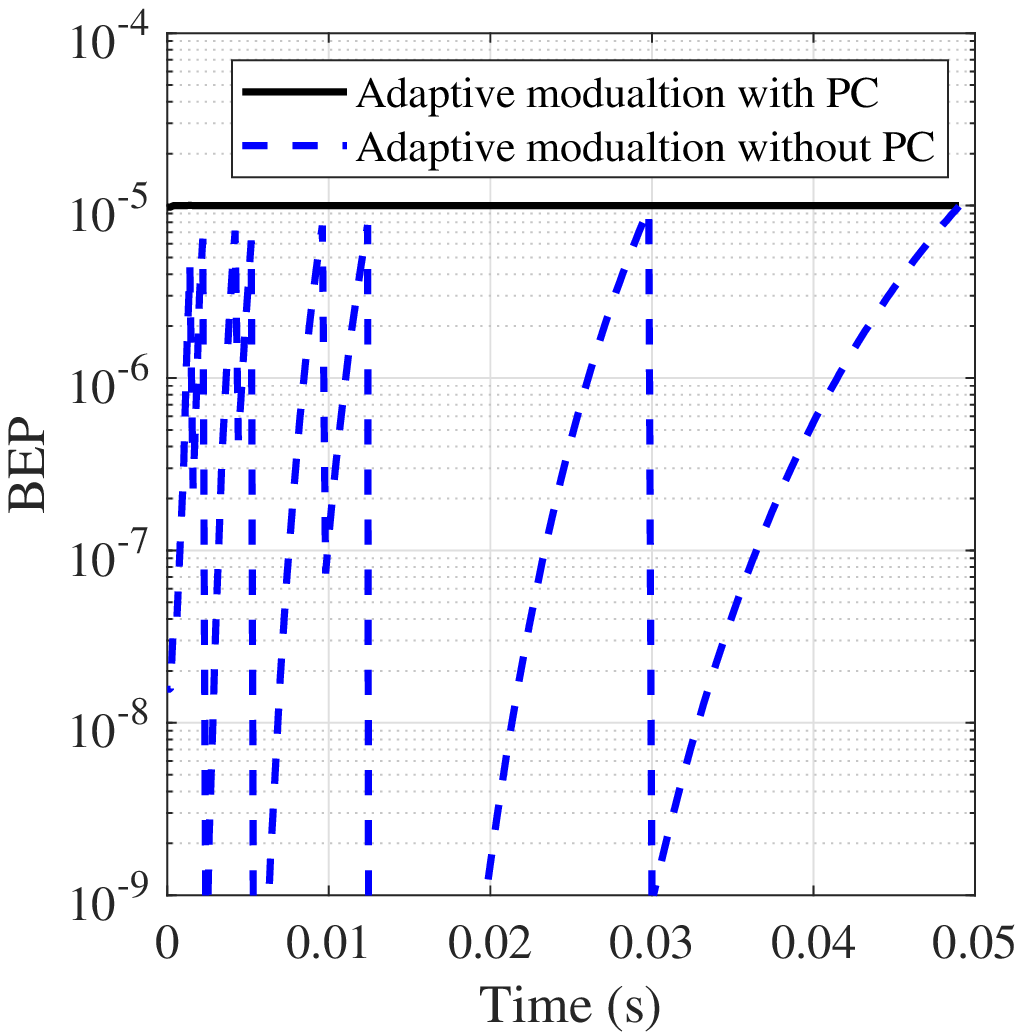}}
		\caption{The performance of power control policy for the adaptive modulation scheme for \textit{M}-ary QAM and BEP threshold at $ 10^{-5} $ under case 1. }
		\label{fig:PCQAM}
	\end{figure}

	According to Fig. \ref{fig:MISO_Time_ABEP_1E5} and \ref{fig:MISO_Time_ABEP_1E52}, the simulation and analytic UUB of  the mm-wave UAV A2G link with the adaptive modulation scheme under imperfect CSI are matched well.
	The simulation is computed based on \eqref{equ:subOR} and the analytic UUB is computed based on \eqref{equ:ABEP_ana} with same estimated channel impulse response. 
	The adaptive modulation scheme decreases the modulation order step by step to keep the instantaneous BEP of the system lower than the BEP threshold for \textit{M}-ary PSK or \textit{M}-ary QAM signals.
	\textit{M}-ary QAM signals have higher modulation order than that of \textit{M}-ary PSK signals because the Euclidean distance of the \textit{M}-ary QAM signals are greater than or equal to that of the \textit{M}-ary PSK signals. 
	
	The instantaneous and average transmission rates with the adaptive modulation scheme under different estimated channel impulse responses are shown in Fig. \ref{fig:RateComparison}.
	The instantaneous transmission rate shows how the adaptive modulation scheme works based on the transmission policy, which is also a supplement of the Fig. \ref{fig:MISO_Time_ABEP_1E5} and \ref{fig:MISO_Time_ABEP_1E52}.
	The instantaneous transmission rate of \textit{M}-ary QAM scheme is more sensitive to $ \mathbf{h}(T_\mathrm{e}) $ than that of \textit{M}-ary PSK scheme.
	The average transmission rate computed by \eqref{eq:Ravedef} increases first and then decreases as transmission time increases.
	
	Fig. \ref{fig:Contour} (a) and (b) show the maximum average transmission rate $ R_\mathrm{ave,max} $ at different value of the BEP threshold and SNR employing \textit{M}-ary PSK or \textit{M}-ary QAM signals. 
	$ R_\mathrm{ave,max}$ increases with an increasing SNR at the same BEP threshold.
	In contrast, $ R_\mathrm{ave,max} $ decreases with the BEP threshold decrease at the same SNR.  
	Since the maximum achievable transmission rate of \textit{M}-ary QAM is higher than that of \textit{M}-ary PSK, \textit{M}-ary QAM have a higher $ R_\mathrm{ave,max} $ than that of \textit{M}-ary PSK at the same BEP threshold and SNR 

	The performance of the adaptive modulation scheme with and without power control for the BEP threshold at $ 10^{-5} $ employing \textit{M}-ary PSK or \textit{M}-ary QAM modulation schemes under case 1 are shown in Fig. \ref{fig:PCPSK} and \ref{fig:PCQAM}.  
	The initial guess root of the Newton-Raphson method for \textit{M}-ary QAM is 30 dB. 
	The transmission power under power control policy can adapt to the different modulation orders for maintaining the instantaneous BEP performance below the BEP threshold. 
	The power control policy makes the  instantaneous BEP approach to the BEP threshold, which facilitates the adaptive modulation scheme to save energy.
	
	For the whole adaptive modulation transmission period, the average transmission power for \textit{M}-ary PSK or \textit{M}-ary QAM modulation scheme with the power control policy at the BEP threshold $ 10^{-5} $ for replication are 21.5 dBm and 21.6 dBm, respectively.
	Therefore, the percentage of power-saving in watts under power control for \textit{M}-ary PSK or \textit{M}-ary QAM modulation scheme are 95.5\% and 95.4\%, respectively.	
	For the optimum adaptive modulation transmission period, the average transmission power for \textit{M}-ary PSK or \textit{M}-ary QAM modulation scheme with the power control policy at the BEP threshold $ 10^{-5} $ for replication are 30.1 dBm and 31.8 dBm, respectively.
	Therefore, the percentage of power-saving in watts under power control for \textit{M}-ary PSK or \textit{M}-ary QAM modulation scheme are 67.7\% and 52.2\%, respectively.
	The \textit{M}-ary PSK wastes more power in the transmission period than \textit{M}-ary QAM and the power control policy is more important for \textit{M}-ary PSK.

	\section{Conclusions}
	 A novel adaptive modulation scheme in conjunction with a power control policy to maximize the average transmission rate and minimize the instantaneous transmission power of the mm-wave UAV A2G link under imperfect CSI is proposed.
	The channel estimation time and the transmission time have an optimized trade-off based on the designed algorithm to achieve the maximum average transmission rate subjects to the BEP threshold.
	The power control policy of the optimum adaptive modulation scheme is presented, which can save 67.7\% and 52.2\% power compared with optimum adaptive modulation without power control, at the BEP threshold $ 10^{-5} $ for \textit{M}-ary PSK or \textit{M}-ary QAM, respectively.
	The proposed work can give a guideline for engineers to design the green mm-wave UAV A2G link for maintaining the BEP performance under the maximum tolerable BEP threshold while maximizing the average transmission rate.
	\appendices
	\section*{Appendix A. Proof of Theorem \ref{Th:ABEP}}
	\label{App:ThABEP}
	According to \eqref{equ:subOR}, the UUB  on the BEP of the sub-optimum detector is given by
	\begin{equation}
		\label{key}
		\mathrm{BEP} \le \sum\limits_{m = 1}^{Nt} {\sum\limits_{\hat m = 1}^{Nt} {\frac{\mathcal{N}_{\left( {m \to \hat m} \right)}\mathrm{P}_{\left( m \to \hat m \right)}}{M{{\log }_2}\left( M \right)}} } .
	\end{equation}
	 According to \cite{Li2019CL,Cho2002TComm}, the PEP is defined by  
	\begin{equation}
		\label{key}
		\begin{aligned}
			\mathrm{P}_{\left(m \to \hat m \right)}	& \buildrel \Delta \over = \mathrm{P} \left( {{\left\| \mathbf{y}\left( t \right) - \sqrt{\gamma} \mathbf{h}\left( T_\mathrm{e} \right)C\left( T_\mathrm{e},t\right) s_m \right\|}^2 > {\left\| \mathbf{y}\left( t \right)- 	\sqrt{\gamma} \mathbf{h}\left( T_\mathrm{e} \right)C\left( T_\mathrm{e},t\right) s_{\hat m} \right\|}^2} \right) \\
			&=  \mathrm{Q}\left( \sqrt { \frac{{\left\|\sqrt{\gamma} \mathbf{h}\left( T_\mathrm{e} \right)C\left( T_\mathrm{e},t\right)  \right\|}^2{\left| s_m - s_{\hat m} \right|}^2}{2{\left| \sqrt {\gamma \left( 1 -C\left( T_\mathrm{e},t\right) ^2 \right){\left| s_m \right|}^2 + 1} \right|}^2} }\right),
		\end{aligned}
	\end{equation}
	where $ \mathrm{Q}\left(x\right) = \frac{1}{\pi }\int\limits_0^{\frac{\pi }{2}} {e^{\left(  - \frac{x^2}{2 \sin ^2\theta} \right)}{\mathrm{d}}\theta } $. 
	\section*{Appendix B. The Table of $ \mathbf{h}(T_{\mathrm{e}}) $ Value in Monte-Carlo Simulation}
	
	\begin{table}[H]
		\centering
		\caption{$ \mathbf{h}(T_{\mathrm{e}}) $ Value in the Monte-Carlo Simulations}
		\begin{tabular}{|c|c|c|}
			\hline
			&Case 1&Case 2\\
			\hline
			$ h_1 $&1.5511+ 0.1561i&0.9578+2.0563i\\
			\hline
			$ h_2 $&-0.4685 + 0.3031i &-0.7581+0.5835i\\
			\hline
			$ h_3 $&0.8435 - 0.3901i&0.6795+0.9751i\\
			\hline
			$ h_4 $&1.2329 + 0.4709i &0.0877-0.7482i\\
			\hline
			$ h_5 $&-0.4971 - 1.1352i &1.0159-0.3314i\\
			\hline
			$ h_6 $&0.1728 + 0.2262i &-1.3866-0.1927i\\
			\hline
			$ h_7 $&0.1781 - 0.4837i &-0.1398+0.7767i\\
			\hline
			$ h_8 $&-0.5205 + 0.6567i&-0.8541-0.1965i\\
			\hline
		\end{tabular}
		\label{table:2}
	\end{table}

	\section*{Appendix C. Proof of Lemma \ref{le:QAM_SNR}}
	According to \eqref{equ:QAM_app}, we can obtain
	\begin{equation}\label{equ:QAM_app_equal}
		f(\gamma_\mathrm{re})=\ln(\beta_{\mathrm{th}})- \ln\left( \sum_{m=1}^{M}{
			\sum_{\hat{m}=1}^{M}{
			\frac{\mathcal{N}_{\left( 	m \to \hat m \right)}\mathrm{Q}\left( \sqrt{\frac{\Lambda \gamma_\mathrm{re}}{\psi \gamma_\mathrm{re}+2}}\right)  }{M{\log }_2\left( M \right)}}
		}\right) =0,
	\end{equation}
	where $ \mathrm{Q}\left(x\right) = \int\limits_x^{+\infty} {\frac{1}{\sqrt{2\pi}}e^{\left(  - \frac{w^2}{2 }\right)} {\mathrm{d}}w  }$.
	
	According to Newton–Raphson method, the approximation root can computed as
	\begin{equation}\label{equ:Newton}
		\ln (\gamma_\mathrm{re+1}) = \ln(\gamma_\mathrm{re})-\frac{f(\gamma_\mathrm{re})}{f^{'}(\gamma_\mathrm{re})}.
	\end{equation} 
	We take the derivation of \eqref{equ:QAM_app_equal} with respect to $ \gamma_\mathrm{re} $ and obtain 
			\begin{equation}\label{equ:QAM_SNR_1}
		\ln(\gamma_\mathrm{re+1})=\ln(\gamma_\mathrm{re})-u_m\frac{\ln(\beta_{th})-\ln(u_m)}{v_m}.
	\end{equation}
	After reforming \eqref{equ:QAM_SNR_1}, we obtain \eqref{equ:QAM_SNR}.
	\bibliographystyle{IEEEtran}
	\bibliography{C:/References/References.bib}
\end{document}